\documentclass[%
aip,
 pof,
 onecolumn,
 notitlepage,
 amsmath,amssymb,
 reprint,
 onecolumn
]{revtex4-1}

\usepackage{graphicx}

\begin{document}

\title{Stabilising falling liquid film flows using feedback control}
\author{Alice B. Thompson}
\email{alice.thompson1@imperial.ac.uk}
\author{
Susana N. Gomes}
\author{
Grigorios A. Pavliotis}
\author{
Demetrios T. Papageorgiou}
\affiliation{ 
Department of Mathematics, Imperial College London, London, SW7 2AZ, UK.}

\date{\today}

\begin{abstract}
Falling liquid films become unstable due to inertial effects when the fluid layer is sufficiently thick or the slope sufficiently steep. 
This free surface flow of a single fluid layer has industrial applications including coating and heat transfer, which benefit from smooth and wavy interfaces, respectively. 
Here we discuss how the dynamics of the system are altered by feedback controls based on observations of the interface height, and supplied to the system via the perpendicular injection and suction of fluid through the wall. In this study, we model the system using both Benney and weighted-residual models that account for the fluid injection through the wall.
We find that feedback using injection and suction is a remarkably effective control mechanism: the controls can be used to drive the system towards arbitrary steady states and travelling waves, and the qualitative effects are independent of the details of the flow modelling. 
Furthermore, we show that the system can still be successfully controlled when the feedback is applied via a set of localised actuators and only a small number of system observations are available, and that this is possible using both static (where the controls are based on only the most recent set of observations) and dynamic (where the controls are based on an approximation of the system which evolves over time) control schemes.
This study thus provides a solid theoretical foundation for future experimental realisations of the control of falling liquid films.
\end{abstract}

\maketitle

\section{Introduction}

The flow of a thin liquid film down an inclined planar wall is a classical problem in fluid mechanics. 
The flow becomes unstable when the Reynolds number, defined on the undisturbed interface flow speed, is above a critical value which depends on the inclination angle; the flow is stable when the layer is sufficiently thin. After the onset of instability, the system initially exhibits two-dimensional (2-D) waves that propagate down the slope, followed by more complicated behaviour that can eventually lead to three-dimensional (3-D) spatiotemporal chaos. 
The development of thin film models for this system, and the behaviour exhibited therein, has recently been reviewed by \citet{Craster_review, Kalliadasis}.

In addition to acting as a paradigm for understanding transitions between different types of dynamical behaviour, the flow of thin films has a broad range of industrial applications. We note particularly coating flows \citep{Weinstein_Ruschak}, where a uniform coating of a flat or shaped substrate is desired, and heat and mass transfer, which is typically enhanced by mixing associated with interfacial waves \citep{Kalliadasis}. The film dynamics and stability can be influenced by feedback control, which may be able to delay the onset of instability, and in an ideal situation we would like to be able to drive the system into the full range of regimes.

Much of the thin-film literature focuses on the additional instabilities and flow modes that can occur in flows with heating and cooling \citep{Bankoff}, or on flows over non-uniform topography \citep{Pozrikidis, Gaskell_topography, Inclined_topography, Heining2009, Pollak2013, Schorner2015}; both have direct applications in heat exchangers. Thermal effects and topography are often combined with each other \citep{Nonuniform_heating, Blyth_Bassom} or with electric fields \citep{Gharraei_etal, Tseluiko_et_al_2008a, Tseluiko_et_al_2008b, Waves_on_electrified_thin_films, Veremeieiev_etal}. Other physical mechanisms that have been investigated within the context of thin-film flow down inclined planes include chemical coatings or microstructure to induce effective slip \citep{Kalpathy_etal}, surfactants \citep{Blyth_Pozrikidis_surfactant}, porous \citep{Thiele, Ogden_etal} or deformable \citep{Gaurav_Shankar} walls, explicit injection/suction through the planar wall \citep{Thompson_steady_suction} and magnetic fields \citep{Amaouche_etal}.
For flat homogeneous walls, a steady uniform flow solution exists known as the Nusselt solution, which can of course be unstable.
In contrast, imposed heterogeneity can be used to create patterned states with different stability properties to the corresponding homogeneous system.
However, systems in which the controls are able to actively respond to the instantaneous system state, rather than being pre-determined (open-loop control), have a much stronger effect on stability. In this paper, we consider such closed-loop (or feedback) control, which will be imposed to the system by suction and injection through the wall. 

\citet{Thompson_steady_suction} used long wave models to study the effect of imposed, steady, spatially-periodic suction/injection on thin-film flow down an inclined plane. They found that the imposed suction always leads to non-uniform states, enables a non-trivial bifurcation structure and complicated time-dependent behaviour, and significantly alters the trajectories of particles in the flow, but has a relatively small effect on flow stability. 
Fluid injection through slots has also been considered theoretically for its effect on spreading films \citep{Momoniat2010}; suction leads to ridges on the free surface, and injection to indentations, but there is no steady state as the total mass is not conserved.
Injection has some similarities to flow over a porous wall, which tends to wick fluid into narrow pores; this flow is particularly relevant to the printing of ink onto paper, for which substrate porosity affects the lifetime and spreading of drops.
\citet{DavisHocking} considered flow of thin drops and films with wetting fronts along a porous substrate that is wetted by the fluid, and is initially dry, and found that for both films and drops, the fluid is eventually drawn entirely into the substrate.
\citet{Thiele} used a Benney equation to study flow over a heated, fluid-filled, inclined porous substrate, bounded below by a solid wall so that the total mass of the liquid film is conserved. They found that the addition of a porous substrate typically has a small destabilising effect on the uniform film state, and in the nonlinear regime the film develops drops and ridges which slide down the plane.
Film flows with point and slot-shaped sources have also been studied as a model for lava spreading \citep{Schwartz_lava}, though the fluid does not form a continuous film, instead containing several wetting fronts.

Feedback control requires observations of at least some components of the system state, and we will build our control strategies around observations of the film height.
\citet{Liu_Gollub_1993} investigated experimentally the dynamics of thin films within the context of the onset of chaos; they used a fluorescence imaging process to measure the two-dimensional film thickness in real time, and also used laser beam deflection to obtain local measurements of the interface slope.
\citet{Bontozoglou_wavy_experiments} examined the flow of thin films over a wavy wall, and used interferometry calibrated against needle-point measurements to obtain the interface height.
\citet{Heining2013} showed that the free surface shape and topography profile can be obtained from measurements of the surface velocity, and implemented this both in Navier--Stokes simulations and experiments.
\citet{Schorner2015} used experiments with visualization by laser reflection to study the effect of differently shaped topographical configurations with the same basic amplitude and wavelength on the flow down an inclined plane; they were able to infer the streamwise growth rate of small-amplitude perturbations by comparing the magnitude of interfacial fluctuations at two streamwise locations.

Thin film flow is often studied using reduced-dimensional models, which differ most fundamentally in the manner in which inertial effects are incorporated.
 Here we use two different first-order long-wave models: the Benney equation and the weighted-residual (WR) equation, which were extended by \citet{Thompson_steady_suction} to include the effect of suction and injection. 
  These two long-wave models are identical at zero Reynolds number, and both agree with the Navier--Stokes system regarding the critical Reynolds number for the onset of instability.
 However, the structures of these models differ significantly, in particular the number of degrees of freedom. The Benney equation is a single evolution equation for the interface height $h(x,t)$, while the weighted-residual model comprises coupled equations for $h(x,t)$, and the independently-evolving down-slope flux $q(x,t)$, and of course the Navier--Stokes equations at finite Reynolds number allow evolution of $h(x,t)$ together with evolution of the flow velocity at every point within the fluid.
 The robustness of control strategies to changes in the model is one of the major themes of this paper;
 we seek to understand what features of the system state need to be measured to deliver effective control, and whether the control system can be designed without needing detailed knowledge of the system state and underlying dynamics.

Feedback control systems consist of a set of control actuators and response functions \citep{Zabczyk1992}; for linear controls the response is a linear function of the deviation of the observed state from the desired state. 
An appropriate linear function is constructed based on hypotheses regarding the dynamics of the uncontrolled system and its response to the control actuators.
In the simplest scenario, the controls are distributed along the domain. In more realistic scenarios, the control actuation is only possible at a finite number of points in the domain, and observations of the current state are not available everywhere. Feedback control theory is useful in both cases, and standard tools for tackling these problems are discussed by \citet{Zabczyk1992}.

Theoretical applications of feedback control for thin films include thermal perturbations in liquids spreading over a solid substrate to suppress the contact line instability \citep{Grigoriev} and point actuated suction/injection to suppress waves in weakly nonlinear models of thin films \citep{Christofides1998} or to enhance such wavy behaviour \citep{Susana_long}. When suppressing waves, again in weakly nonlinear systems, \citet{Armaou2000} prove that stabilisation can be achieved by using a finite number of observations, either by static or dynamic output feedback control (see Sec.~\ref{sec:Evolving_controls} and appendix \ref{appB}), while \citet{Armaou2000a} use nonlinear feedback controls. Efforts have also been made to optimise the placement of actuators and sensors to suppress \citep{Lou2003} or enhance \citep{Susana_long} waves  on a thin film.
Feedback control strategies have been implemented for the two-dimensional Navier--Stokes equations in the context of data assimilation \citep{Farhat2015}, in which controls applied towards known observations are used to overcome incomplete knowledge of the initial state in the forecasting of hurricanes and typhoons.

Both of the long wave models studied here reduce to the forced Kuramoto-Sivashinsky (KS) equation under a weakly nonlinear analysis, and form part of a rational hierarchy of models which lead to the KS equation \citep[see e.g.][]{ShlangSiv, SivMich}.
The KS equation retains many essential features of the Benney equation, in particular nonlinearity, energy production and dissipation. However, it is a much simpler system, which moreover has the property of global existence of solutions \citep{Tadmor1986}, a global attractor \citep{Constantin}, and sharp bounds on its solutions and derivatives \citep{Otto2009}. These global properties make the KS equation amenable to analysis, and various control schemes have been developed and applied for the KS equation, e.g. \citet{Susana_long}. With the inclusion of suitable linear feedback controls towards the flat solution, the bounds on the solutions make it possible to prove that the $L^2$-norm is a Lyapunov function for the dynamical system corresponding to the KS equation, and hence the controlled system is nonlinearly stable. Furthermore the existence of bounds on the solutions can be used to prove the existence of optimal controls. When stabilising nontrivial steady states or travelling waves, the boundedness of these nontrivial solutions play a crucial role in the proof of existence of a Lyapunov function and also influence the choice of the controls.

The difficulty in applying the KS controls to thin-film systems is twofold: the nonlinearities are more complicated, and so there is no global existence theory; in fact the Benney model can lead to unbounded behaviour \citep{Pumir, Kalliadasis}, and secondly, the structure of the weighted-residual and Navier--Stokes models, when applied at finite wavelength, is significantly different to that of the KS equation. It is reasonable to suspect that a control strategy carefully optimised for one model may be ineffective in another, and so we focus here on the use of relatively simple control schemes, and investigate their robustness to variations in the model details.

In this paper, we consider the effect of feedback control, applied via a suction boundary condition and based on observations of the interface height, on the dynamics and stability of a thin layer of fluid flowing down an inclined plane. We use two different long-wave models, described in Sec.~\ref{sec:Governing_equations}, to model the system, and also compare certain stability results to those for the Navier--Stokes equations. 
In Sec.~\ref{sec:Control_to_uniform}, we show that a proportional control scheme, in which fluid is injected at each streamwise location in proportion to the observed deviation of the interface height at that location from uniform,
has a stabilising effect on nearly-uniform flow in all three models. We compute the critical control magnitude required to stabilise the uniform state to perturbations of all wavelengths, finding that the critical Reynolds number for the onset of propagating waves can be increased significantly by using the proportional control scheme.
For the control system to be physically realisable, we must relax the requirement that the system state is known at every position, and that the suction applied can take an arbitrary shape. Therefore, in Sec.~\ref{sec:Actuators_and_observations}, we discuss control strategies for the case where feedback is applied through a discrete set of localised actuators and the system is similarly observed at a small number of locations in the domain. We discuss both control schemes based on static observations (where the control amplitudes are calculated from only the most recent set of observations) and dynamic observations (where the controls are calculated using an approximation to the system state which is built up over time).  
In Sec.~\ref{sec:Non_uniform}, we discuss linear stability and nonlinear behaviour when controlling to non-uniform travelling waves and non-uniform steady states.
We note that the property of a given non-uniform state being an exact solution of the equations is model dependent and therefore can never be perfectly satisfied. In order to test the robustness of the control scheme to variations in the model, we consider controlling to non-uniform interface shapes with only very crude prior knowledge of the governing equations. We find that in this case, the controls lead to equilibrium states which converge towards the desired system as the control amplitude is increased, and that these equilibrium states are stable when large-amplitude controls are applied.
Our conclusions are presented in Sec.~\ref{sec:Conclusion}.

\section{Governing equations}
\label{sec:Governing_equations}

\begin{figure}
\centering
 \includegraphics[width=0.7\textwidth]{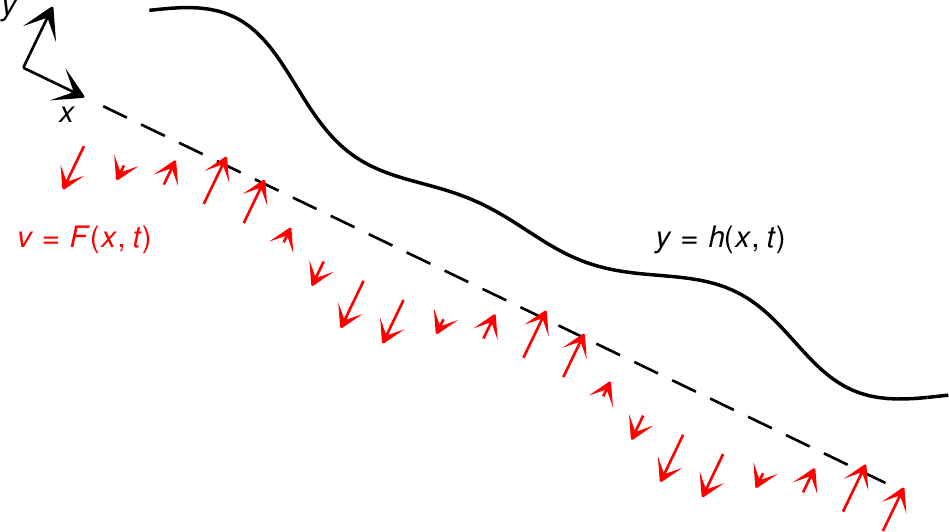}
 \caption{Sketch of flow domain showing coordinate system. We consider a fluid layer, with mean height $h_s$, bounded on $y=0$ by a rigid wall inclined at an angle $\theta$ to the horizontal, and at $y=h(x)$ by a free surface. Fluid is injected through the wall, with velocity $v = F(x,t)$ which changes in time in response to fluctuations of the free surface. \label{fig:Inclined_plane_figure}}
\end{figure}

We consider a thin layer of fluid, with mean thickness $h_s$, flowing down a plane inclined at an angle $\theta$ to the horizontal. We adopt a coordinate system such that $x$ is the down-slope coordinate, and $y$ is the perpendicular distance from the wall, as shown in Fig. \ref{fig:Inclined_plane_figure}. The upper interface of the fluid is a free surface, located at $y=h(x,t)$. The lower boundary of the fluid is a rigid wall, through which fluid may be injected or removed. 

The two-dimensional (2-D) Navier--Stokes equations admit a solution which is uniform in the streamwise direction, known as the Nusselt solution \citep{Nusselt}, for which the surface velocity is $U_s = \rho g h_s^2 \sin\theta/(2\mu)$, where $\rho$ is the fluid density, $g$ the acceleration due to gravity and $\mu$ the dynamic viscosity of the fluid.
We non-dimensionalise the problem based on the length scale $h_s$ and the velocity scale $U_s$, and define the Reynolds number $R$ and the capillary number $C$ based on the velocity $U_s$:
\begin{equation}
 R = \frac{\rho h_s U_s}{\mu}, \quad C = \frac{\mu U_s}{\gamma},
\end{equation}
where $\gamma$ is the coefficient of surface tension at the air-fluid interface. Subsequent equations are all dimensionless.

\subsection{Navier--Stokes equations}

We wish to solve the 2-D Navier--Stokes equations, with velocity $\mathbf{u}(x,y,t)=(u,v)$, and fluid pressure $p(x,y,t)$. The momentum equation and continuity equations are
 \begin{eqnarray}
 \label{NS_xmom}
R\left(u_t + u u_x + vu_y\right) 
= -  p_x
 +2+u_{xx}+ u_{yy},
\end{eqnarray}
\begin{eqnarray}
\label{NS_ymom}
R \left(v_t + u v_x + v v_y\right) = 
-  p_y
-2 \cot\theta 
+v_{xx}+v_{yy}
\end{eqnarray}
and
\begin{eqnarray}
\label{NS_mass}
 u_x + v_y=0.
\end{eqnarray}

The boundary conditions at the wall are given by
\begin{equation}
\label{NS_wall}
 u =0, \quad  v=F(x,t).
\end{equation}
Here the function $F(x,t)$ represents the injection velocity normal to the wall, $y=0$. Note that we assume that the injection of fluid does not affect the no-slip boundary condition on the wall.
At the interface, $y=h(x,t)$, the tangential and normal components of the dynamic stress balance condition yield
\begin{eqnarray}
\left (v_x+u_y\right)\left (1-h_x^2\right) + 2h_x \left(v_y-u_x\right)&=&0,\\
\displaystyle  p - p_a
 - \frac{2}{1+ h_x^2} 
 \left(
v_y
+ u_x h_x^2   - h_x\left(v_x+u_y\right)\right) 
 &=& -\frac{1}{C}\frac{h_{xx}}
 {
 (1+h_x^2)^{3/2}
 },
 \label{non-scaled-end}
\end{eqnarray}
where $p_a$ is the atmospheric pressure, assumed constant.
The system is closed by the kinematic boundary condition at the free surface
\begin{equation}
\label{NS_kin}
 h_t = v - u h_x.
\end{equation}

Defining the down slope flux $q$
\begin{equation}
 q(x,t) = \int_0^h u(x,y,t)\, \mathrm{d}y,
\end{equation}
integrating \eqref{NS_mass}, and applying the boundary conditions \eqref{NS_wall} and \eqref{NS_kin}, yields the mass conservation equation in terms of $q$:
\begin{equation}
\label{mass_governing}
 h_t - F(x,t) + q_x=0.
\end{equation}

The flow is modelled either by the 2-D Navier--Stokes equations, or by one of two reduced-dimension long-wave models, which are derived according to either the Benney \citep{Benney} or weighted-residual \citep{Ruyer_Quil} methodology in order to approximate the flow in a long wave limit. To achieve this, we define new variables
\begin{equation}
\label{long_wave_scaling}
 X = \delta x, \quad T = \delta t, \quad v = \delta w, \quad C = \delta^2 \widehat{C}, \quad F = \delta f
\end{equation}
and seek a solution for small $\delta$.
The extension of these two long wave models to include the effects of suction/injection is discussed by \citet{Thompson_steady_suction}, and so the relevant governing equations are stated without derivation here. 

\subsection{Benney system}

In the Benney model, the flux $q$ is slaved to the interface height $h$, and up to first order in $\delta$, including the cross-flow effects induced by $F$, the flux is given by \citep{Thompson_steady_suction}:
\begin{equation}
 q(X,T) =  \frac{2h^3}{3}
 + \delta \left[ \frac{h^3}{3}\left( 
 -2 h_X\cot\theta + \frac{h_{XXX}}{\widehat{C}}\right)
+
R \left( 
\frac{8h^6h_X }{15}
-\frac{  2h^4 f}{3}
\right)\right].
\end{equation}
We then recast the equation in terms of the original variables to obtain
\begin{equation}
\label{eq:Benney}
 q(x,t) =  \frac{h^3}{3}\left( 2
 -2 h_x\cot\theta + \frac{h_{xxx}}{C}\right)
+
R \left( 
\frac{8h^6h_x }{15}
-\frac{  2h^4 F}{3}
\right) = Z(h,F).
\end{equation}
The coupling of \eqref{eq:Benney} to \eqref{mass_governing} yields a closed system for the evolution of the interface height $h(x,t)$.

We note that the appearance of terms involving $F$ in \eqref{eq:Benney} is a consequence of the choice of $F$ with respect to the long wave scaling. By supposing $F$ to be an order smaller with respect to $\delta$ in the long wave expansion \eqref{long_wave_scaling}, we can replace \eqref{eq:Benney} with the simpler version:
\begin{equation}
\label{eq:Benney_simplified}
 q(x,t) =  \frac{h^3}{3}\left( 2
 -2 h_x\cot\theta + \frac{h_{xxx}}{C}\right)
+
\frac{8Rh^6h_x }{15}.
\end{equation}
In this limit, the only effect of $F$ on the system dynamics is via the mass conservation equation \eqref{mass_governing}.

\subsection{Weighted-residual system}

Alternatively, following the weighted-residual methodology developed by \citet{Ruyer_Quil}, the flux $q$ gains its own evolution equation, so that time derivatives of both $h$ and $q$ appear in the equations. After substituting \eqref{long_wave_scaling} into the governing equations and retaining terms up to and including $O(\delta)$, the evolution equation for $q$ is \citep{Thompson_steady_suction} 
\begin{equation}
\label{eq:WR_delta}
 \frac{2 \delta R h^2}{5} \frac{\partial q}{\partial T} + q = \frac{2h^3}{3} +\delta\left[\frac{h^3}{3}\left(
 -2 h_X\cot\theta + \frac{h_{XXX}}{\widehat{C}}\right)
 + R \left( \frac{18q^2 h_X}{35} - \frac{34 h q q_X}{35}+ \frac{ hqf}{5}\right)\right].
\end{equation}
In the original variables, \eqref{eq:WR_delta} becomes
\begin{equation}
\label{eq:WR}
 \frac{2}{5}R h^2 q_t + q = \frac{h^3}{3}\left( 2
 -2 h_x\cot\theta + \frac{h_{xxx}}{C}\right)
 + R \left( \frac{18q^2 h_x}{35} - \frac{34 h q q_x}{35}+ \frac{ hqF}{5}\right) = Z(h,q,F),
\end{equation}
which when coupled to \eqref{mass_governing} yields a closed system, for $h(x,t)$ and $q(x,t)$. Initial conditions are required for both $h$ and $q$. The Benney and weighted-residual models are identical when $R=0$, and can be shown to agree at $O(1)$ and $O(\delta)$ in the long-wave limit \citep{Thompson_steady_suction}.

\subsection{Choice of controls}

The focus of this paper is on the application of suction as a linear control mechanism in response to observations of the interface height. We begin in Sec.~\ref{sec:Control_to_uniform} by considering the case of controlling towards the uniform Nusselt state, based only on observations of $h$. To achieve this, we set
\begin{equation}
\label{uniform_control_scheme}
 F(x,t) = -\alpha[h(x,t)-1],
\end{equation}
where $\alpha$ is a real constant to be chosen; in most cases we find that the uniform state becomes increasingly stable for large positive $\alpha$. Note that if $h=1$ everywhere, then the controls have zero magnitude.

The control scheme \eqref{uniform_control_scheme} requires perfect knowledge of the instantaneous interface shape $h(x,t)$, and the ability to impose any continuous $F(x,t)$. In practice, we expect neither of these assumptions to hold. Instead, fluid is injected via a number of localised actuators, or slots, in the substrate, and interface observations are available at a small number of locations in the flow domain. In Sec.~\ref{sec:Actuators_and_observations}, we investigate control schemes based on point actuators and localised observers, with both static observations  and dynamic observers, and the appropriate form of the injection profile $F(x,t)$ will be discussed when necessary.

In Sec.~\ref{sec:Non_uniform}, we consider controlling towards either nonuniform travelling waves of permanent form or nonuniform steady states. Travelling waves can be written as $h = H(\zeta)$, where $\zeta = x-Ut$ and $U$ is the constant propagation speed. By direct analogy to \eqref{uniform_control_scheme}, we set
\begin{equation}
 F(\zeta,t) = -\alpha [h(\zeta,t)-H(\zeta)].
\end{equation}
We note that if $h(x,t) = H(x-Ut)$ for all time, then $F=0$, so that the travelling wave $h=H(x-Ut)$, is also a solution of the controlled equations.
Nonuniform steady interface shapes $H(x)$ are not steady states of the equations when $F=0$, but \citet{Thompson_steady_suction} showed that imposing a steady suction component $S(x)$ enables non-uniform steady states. Combining with linear control, we obtain
\begin{equation}
 F(x,t) = -\alpha[h(x,t)-H(x)] + S(x).
\end{equation}
For non-uniform states, the calculation of $S(x)$ to obtain an exact steady state, or of the travelling wave solution $H(\zeta)$, require detailed knowledge of the governing equations. For example, these states differ even between the Benney and weighted-residual models, let alone the Navier--Stokes equations. In Sec.~\ref{sec:Imperfect_model_knowledge}, we consider the robustness of our control schemes when the model details are not well known; we do so by controlling towards a finite-amplitude non-uniform state $H(x)$, but setting $S(x)=0$, so that the target state is not a steady solution. As a result, the control parameter $\alpha$ has a role to play in setting both the shape of any steady states obtained, as well as their stability.

\subsection{Numerical calculations for linear stability of non-uniform solutions}
\label{sec:Eigenvalue_problems}

For a translationally invariant system, as occurs for distributed controls towards a uniform film state, the linear stability of the uniform film state can be calculated via a normal mode analysis, and this will be pursued for the Benney, weighted-residual and Navier--Stokes systems in section \ref{sec:Control_to_uniform}.  
However, if the base state for the stability analysis is not uniform, or the feedback control system has localised actuators or observers, then the system is no longer translationally invariant, and so the eigenmodes of the system are no longer normal modes. In that case, we can compute the discretised eigenmodes of the system by formulation and numerical solution of a generalised eigenvalue problem for linear stability, as described below.

We consider the evolution of a small perturbation $\hat{h}$:
\begin{equation}
 h = H(x) + \epsilon \hat{h}e^{\lambda t}, \quad q = Q(x) + \epsilon \hat{q} e^{\lambda t}, \quad F = S(x)- \epsilon e^{\lambda t} \alpha \hat{h}.
\end{equation}
We recall the Benney equation:
\begin{equation}
 h_t + q_x - F=0, \quad q = Z(h,F),
\end{equation}
where $Z(h,F)$ is defined in \eqref{eq:Benney}, and expand for small $\epsilon$.
The equations at $O(1)$ in $\epsilon$ must be satisfied by the base state $H(x)$, $Q(x)$, $S(x)$.
At $O(\epsilon)$, we obtain a generalised eigenvalue problem for $\hat{h}$, $\hat{q}$ and $\lambda$:
\begin{equation}
\lambda
 \begin{pmatrix}
  I & 0 \\
  0 & 0
 \end{pmatrix}
  \begin{pmatrix}
  \hat{h} \\
  \hat{q}
 \end{pmatrix} = 
  \begin{pmatrix}
-\alpha I & -\partial_x \\
 Z_h - Z_F \alpha I & -I
 \end{pmatrix}
  \begin{pmatrix}
  \hat{h} \\
  \hat{q}
 \end{pmatrix}
\end{equation}
where $\partial_x$ is the derivative operator, $I$ is the identity matrix, and 
the blocks $Z_h$ and $Z_F$ are linear operators, for example from \eqref{eq:Benney} we have
\begin{equation}
 Z_h = \left[ H^2\left(2- 2\cot\theta H_x + \frac{ H_{xxx}}{C}\right) + \frac{16R H^5 H_x}{5} - \frac{8H^3 R S}{3}\right] I + \left[-\frac{2 H^3}{3}\cot\theta+ \frac{8R}{15}H^6\right] \partial_x + \frac{H^3}{3C} \partial_{xxx}.
\end{equation}
We can eliminate $\hat{q}$ to obtain a smaller eigenvalue problem for $\hat{h}$ alone:
\begin{equation}\label{eq:small_e_value}
 \lambda \hat{h} = -\partial_x Z_h \hat{h} - [I-\partial_x Z_F]\alpha \hat{h}.
\end{equation}
For the uniform state, $H=1$, $Q=2/3$, and $S=0$, the blocks $Z_h$ and $Z_F$ simplify considerably; in fact, we can calculate the eigenvalues analytically in that case. For non-uniform base states, we calculate the eigenvalues by replacing the derivative operators with pseudo-spectral derivative matrices, and solving the eigenvalue problem numerically
using standard algorithms available in \textsc{Matlab}.

We can also write the flux equation of the weighted-residual system in a similar form:
\begin{equation}
 \frac{2}{5}R h^2 q_t + q  = Z(h, q, F).
\end{equation}
We again obtain a generalised eigenvalue problem for $\hat{h}$, $\hat{q}$ and $\lambda$ in the weighted-residual equations:
\begin{equation}
\lambda
 \begin{pmatrix}
  I & 0 \\
  0 & \frac{2}{5}R H^2 I
 \end{pmatrix}
  \begin{pmatrix}
  \hat{h} \\
  \hat{q}
 \end{pmatrix} = 
  \begin{pmatrix}
  -\alpha I & -\partial_x \\
  Z_h-Z_F \alpha I & Z_q-I
 \end{pmatrix}
  \begin{pmatrix}
  \hat{h} \\
  \hat{q}
 \end{pmatrix},
\end{equation}
where the blocks $Z_h$, $Z_q$ and $Z_F$ can be obtained by differentiating \eqref{eq:WR}.
We note that there are twice as many eigenmodes in the weighted residual equations as in the Benney equation.

In addition to linear stability calculations, we also perform a number of initial value calculations as described by \citet{Thompson_steady_suction}, involving a pseudo-spectral method for spatial discretization, and a fully-implicit, backward finite difference time stepper.

\section{Effect of proportional controls on the stability of a uniform film}
\label{sec:Control_to_uniform}

The uniform film state $h=1$, known as the Nusselt solution, is a steady solution to all three sets of equations (Navier--Stokes, Benney and weighted-residual) in the absence of suction. The base state is
\begin{equation}
\label{uniform_film_solution}
 h = 1, \quad q = 2/3, \quad u = y(2-y), \quad v=0, \quad p = 2(1-y) \cot\theta.
\end{equation}
In 2-D Navier--Stokes \citep{Benjamin, Yih}, Benney \citep{Benney} and weighted-residual models \citep{Ruyer_Quil}, this solution is linearly stable to perturbations of all wavelengths if 
\begin{equation}
 R < R_0 \equiv \frac{5}{4}\cot\theta.
\end{equation}
As $R$ is increased across this threshold, the first perturbations to become unstable are those with infinite wavelength, and in fact the long-wavelength nature of the instability was the physical motivation for the development of long-wave models.

The application of linear controls $F = -\alpha(h-1)$ affects the linear stability of the Nusselt solution. As the system is invariant under translation in $x$, the eigenmodes are proportional to $\exp(ikx)$, and so we write
\begin{equation}
\label{uniform_eigenmode_expansion}
 h = 1 + \epsilon \hat{h} e^{ikx+\lambda t}, \quad q = \frac{2}{3} + \epsilon \hat{q} e^{ikx+\lambda t}
\end{equation}
and seek a solution for $\epsilon \ll 1$. We aim to compute $\lambda(k)$; solutions are stable to perturbations of all wavelengths if the real part of $\lambda$, $\Re(\lambda)$, is negative for all real $k$. In what follows we calculate $\lambda$ for each of the models including feedback control in order to establish that the constant $\alpha$ can be chosen to stabilise the uniform flow \eqref{uniform_film_solution}. In the case of Benney and weighted-residual models this can be performed analytically, whereas for the Navier--Stokes equations we compute the eigenvalues numerically.

\subsection{Benney equations}\label{subs:Benney}

The linearised mass conservation equation \eqref{mass_governing} yields
\begin{equation}
\label{uniform_mass_stability}
 \lambda \hat{h} + \alpha \hat{h} + ik \hat{q}=0.
\end{equation}
Substituting \eqref{uniform_eigenmode_expansion} into \eqref{eq:Benney} gives
\begin{equation}
\label{eq:Benney_q_linearised}
 \hat{q} = \left( 2- \frac{2ik\cot\theta}{3} - \frac{ik^3}{3C} + \frac{8ikR}{15} + \frac{2\alpha R}{3}\right)\hat{h},
\end{equation}
and combining \eqref{eq:Benney_q_linearised} with \eqref{uniform_mass_stability} yields a single eigenvalue $\lambda$:
\begin{equation}
\label{Benney_dispersion}
 \lambda = -\alpha\left(1+  \frac{2Rik}{3}\right) - 2ik + \frac{8k^2}{15}\left(R-\frac{5\cot\theta}{4} - \frac{5k^2}{8C}\right).
\end{equation}

Throughout this paper, we will suppose that $\alpha$ is real and independent of $k$, and taking $\alpha>0$ in \eqref{Benney_dispersion} is seen to have a stabilising effect on the Benney system. If $R<R_0$, the Nusselt solution is linearly stable for all real $k$ in the absence of controls, and becomes more so as $\alpha$ increases. However, if $R>R_0$, there is a finite $k$ with maximum growth rate, and it is easy to show that 
\begin{equation}
 \max_k \Re(\lambda) = -\alpha + \frac{16C (R-R_0)^2}{75}.
\end{equation}
Hence we can stabilise the uniform film state against perturbations of all wavelengths by choosing $\alpha > \alpha_B$, where
\begin{equation}
\label{Benney_mu_critical}
\alpha_B =   \frac{16C (R-R_0)^2}{75}.
\end{equation}

\begin{figure}
\includegraphics[width=0.8\textwidth]{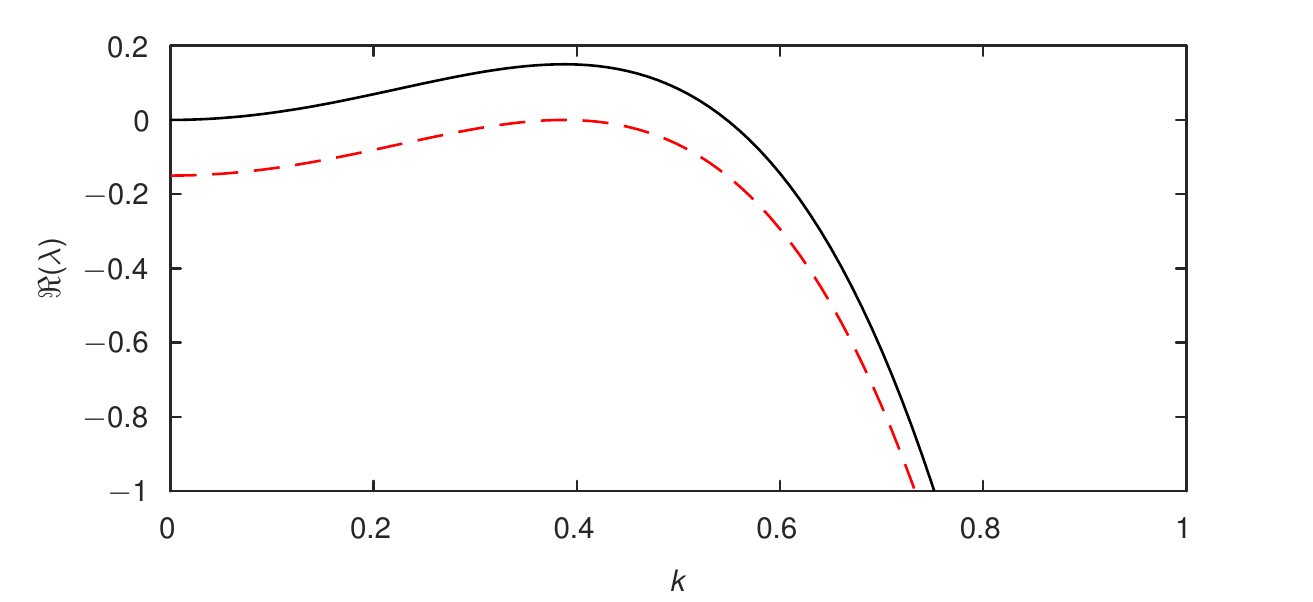}
 \caption{Results for the real part of the Benney eigenvalue $\lambda$ as a function of $k$, for $R=5$, $C=0.05$, $\theta=\pi/4$ and $\alpha=0$ (solid lines), and $\alpha=\alpha_B$ (dashed lines) from \eqref{Benney_mu_critical}. 
 \label{fig:Benney_eigenvalue}}
\end{figure}

The dispersion relation \eqref{Benney_dispersion} is plotted with and without controls in Fig. \ref{fig:Benney_eigenvalue}, for parameters at which the uncontrolled solution is unstable. In the absence of controls, the real part of the $\lambda$ is positive for small $k$, with a finite cutoff wavenumber $k$, above which the real part rapidly becomes increasingly negative. 
Setting $\alpha=\alpha_B$ shifts the real part of the entire spectrum by $-\alpha_B$. This means that perturbations of very small wavenumber decay with a finite growth rate of approximately $-\alpha_B$, rather than having a small positive growth rate in the absence of controls. The maximum growth rate occurs at the same $k$ as in the absence of controls, and for $\alpha=\alpha_B$, this maximum growth rate is exactly zero. 
We can also compare the imaginary part of $\lambda$ (not shown); we find that setting $\alpha=\alpha_B$ slightly increases the magnitude of the imaginary part, and hence the downstream propagation speed of small perturbations is slightly increased.

\subsection{Weighted residual equations}\label{subs:WR}

\begin{figure}
\includegraphics[width=0.8\textwidth]{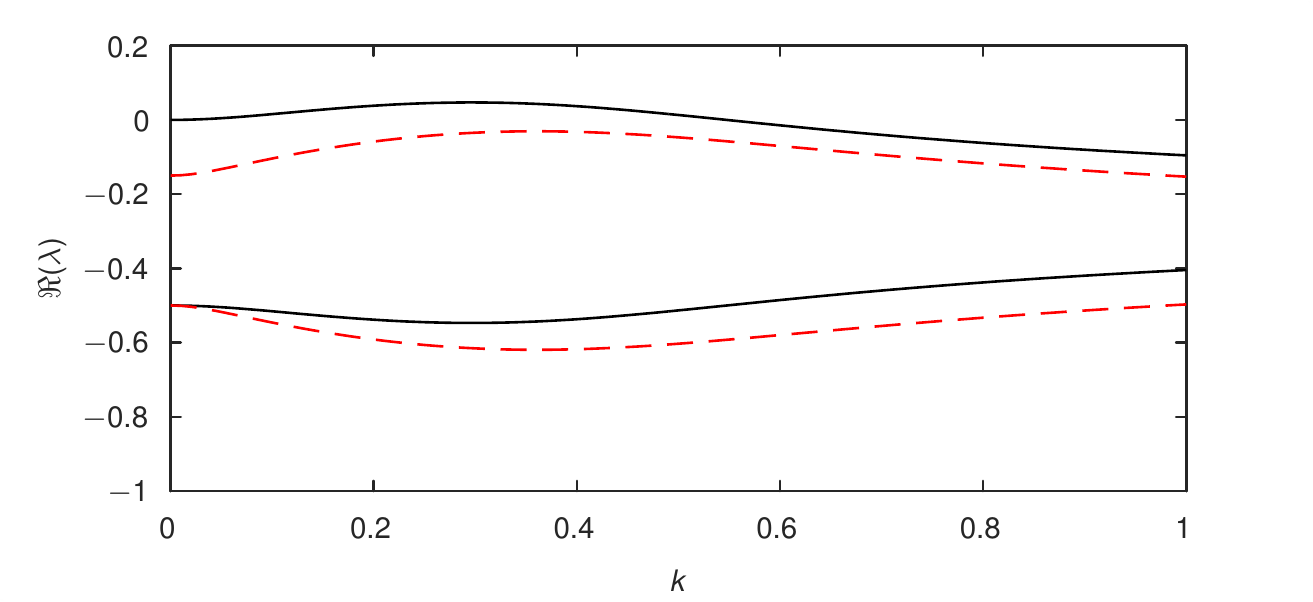}
 \caption{Results for the real part of both eigenvalues $\lambda$ for the weighted-residual model as a function of $k$, for $R=5$, $C=0.05$, $\theta=\pi/4$ and $\alpha=0$ (solid lines), and $\alpha=\alpha_B$ (dashed lines) from \eqref{Benney_mu_critical}. 
 \label{fig:WR_eigenvalues}}
\end{figure}

The linearised version of the weighted residual equations \eqref{eq:WR} yields 
\begin{equation}
\label{eq:WR_q_linearised}
\frac{2\lambda R}{5} \hat{q} + \hat{q} = \left(2 - \frac{2ik\cot\theta}{3} - \frac{ik^3}{3C}  + \frac{8ikR}{35} - \frac{2R\alpha}{15}\right)\hat{h} - \frac{68ikR}{105} \hat{q}.
\end{equation}
We combine \eqref{eq:WR_q_linearised} with \eqref{uniform_mass_stability} to obtain a quadratic equation for $\lambda$:
\begin{equation}
 \label{WR_lambda_eq}
 \frac{2R\lambda^2}{5} + \lambda\left( 1+ \frac{68ikR}{105} + \frac{2\alpha R}{5}\right) + \alpha\left(1 + \frac{18ikR}{35}\right) + 2ik + \frac{8k^2 R_H}{15} - \frac{8k^2 R}{35}=0,
\end{equation}
where
\begin{equation}
 R_H = \frac{5}{4}\cot\theta + \frac{5k^2}{8C} = R_0 + \frac{5k^2}{8C}.
\end{equation}
The characteristic equation
\eqref{WR_lambda_eq} has complex coefficients, and so its two roots for $\lambda$ are not complex conjugates.
We calculate the two roots for $\lambda$ numerically to determine the effect of imposing controls; Fig. \ref{fig:WR_eigenvalues} shows $\lambda$ as a function of $k$, with and without controls. The eigenvalues of the weighted-residual equation display relatively little variation with respect to $k$ in comparison to the Benney results, but the two systems share the same cutoff wavenumber in the absence of feedback controls.
With the addition of feedback controls, we find that positive $\alpha$ decreases the real part of $\lambda$ for both eigenvalues of the weighted-residudal system, with the exception of the most stable eigenmode at $k=0$, which is independent of $\alpha$ (see below). 
Choosing the critical $\alpha=\alpha_B$ for the Benney equation, given by \eqref{Benney_mu_critical}, is more than sufficient to stabilise the uniform state against perturbations of all wavenumbers in the weighted-residual equations.

In the long-wave limit $k\ll 1$, \eqref{WR_lambda_eq} becomes
\begin{equation}
\label{small_k_quadratic}
(\lambda + \alpha) \left( 1 + \frac{2R\lambda}{5}\right)=0
\end{equation}
which has roots at $\lambda=-\alpha$ and $\lambda = -5/(2R)$. Choosing non-zero $\alpha$ affects the stability of the first root, and means that we must choose $\alpha>0$ to obtain a stable solution. The second root is unaffected by $\alpha$, and as a consequence, the maximum real part of $\lambda$ across all $k$ is always greater than $-5/(2R)$, regardless of the value of $\alpha$.

\begin{figure}
 \includegraphics[width=0.8\textwidth]{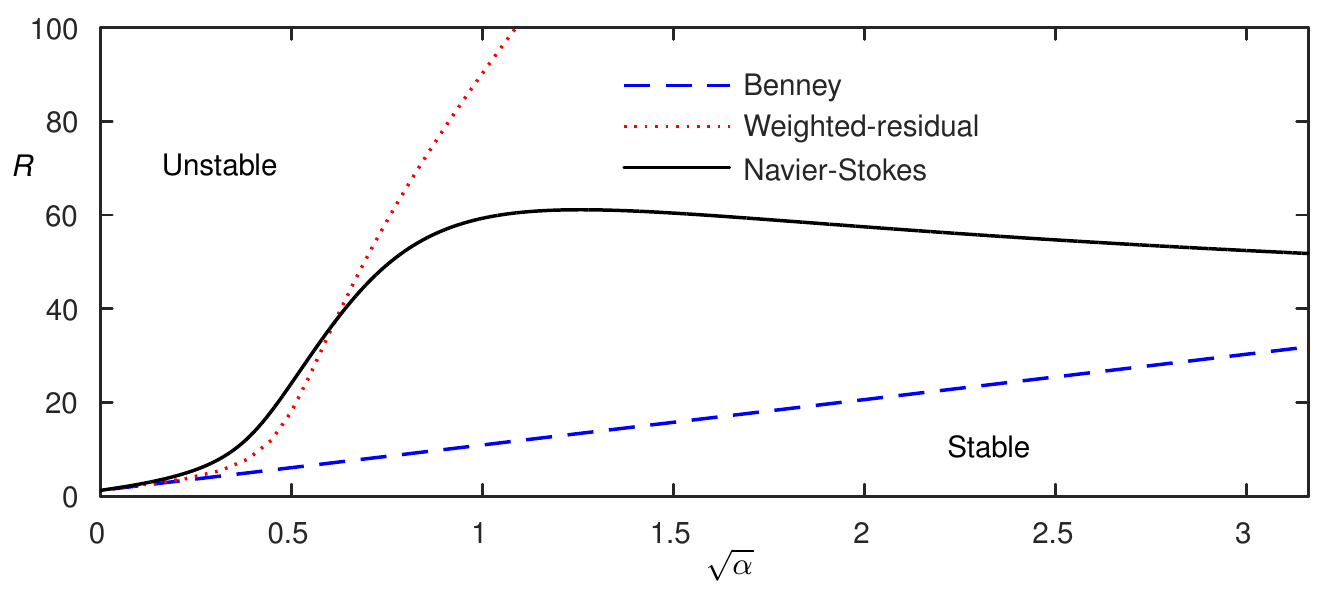}
 \caption{The boundaries for stability to perturbations of all wavelengths, for $\theta=\pi/4$, $C=0.05$. The stable region emanates from the $\sqrt\alpha$ axis.\label{fig:Neutral_stability_map}}
\end{figure}

Although the effect of $\alpha$ on $\lambda$ is more complicated than that for the Benney equations, we can still calculate the critical control amplitude $\alpha$ needed to ensure that $\Re(\lambda) \leq 0$ for all $k$. Perturbations with very large wavenumber are always stabilised by surface tension, so have negative real part. If the uniform state is unstable to perturbations for some $k$, then there is at least one cutoff value of $k$ for which $\Re(\lambda)=0$. 
We therefore investigate the conditions for which there is a purely imaginary root, writing for convenience $\lambda = -2ik\Omega$.
We solve the imaginary part of \eqref{WR_lambda_eq} to obtain
\begin{equation}
 \Omega = \frac{1 + \dfrac{9\alpha R}{35}}{1 + \dfrac{2\alpha R}{5}}.
\end{equation}
Since $\Omega$ is independent of $k$, we can rewrite the real part of \eqref{WR_lambda_eq} as a quadratic equation in $k^2$:
\begin{equation}
\label{quadratic_for_ksquared}
 \frac{k^4}{3C} + k^2 \left( -\frac{8R\Omega^2}{5} + \frac{136R\Omega}{105} - \frac{8R}{35} + \frac{8R_0}{15}\right) + \alpha=0.
\end{equation}
The roots of this equation correspond to wavenumbers where $\Re(\lambda(k))=0$. When $\alpha$ is insufficient to stabilise perturbations of all wavelengths, there are two roots for $k^2$, and one root at the critical value of $\alpha$.
The uniform state is stable to perturbations of all wavelengths if there are no real roots for $k^2$, i.e. when \eqref{quadratic_for_ksquared} has negative determinant.
This condition can be rewritten using the definition of $\Omega$ to obtain that the uniform state is stable if
\begin{equation}
\label{WR_critical_alpha}
\left(
  R\left[\frac{ 
   1
  +\dfrac{71 \alpha R}{245}
  + \dfrac{3 \alpha^2 R^2}{175}
  }{1 + \dfrac{4\alpha R}{5} + \dfrac{4\alpha^2 R^2}{25}}\right]-R_0\right)^2
   <  \frac{75\alpha}{16C}.
\end{equation}
The term in square brackets is monotonically decreasing in $\alpha R$ for $\alpha R>0$.
When $\alpha$ is small, we find 
\begin{equation}
R \approx R_0 + \sqrt\frac{75\alpha}{16C},
\end{equation}
which is exactly the Benney result.
At large $\alpha$, 
\begin{equation}
R \approx \frac{28}{3}\left( R_0 + \sqrt\frac{75\alpha}{16C} \right)
\end{equation}
and so the maximum $R$ for which the uniform solution is stable at fixed $\alpha$
is increased by a factor of approximately 10 according to the weighted-residual model. The stability boundaries for the Benney and weighted residual results are given by \eqref{Benney_mu_critical} and \eqref{WR_critical_alpha}, and are plotted in the $\sqrt{\alpha}$-$R$ plane in Fig.~\ref{fig:Neutral_stability_map}, together with the corresponding Navier--Stokes results as discussed below; it appears that the stable region of the Benney equation is always a subset of the stable region according to the weighted-residual equation, so the critical $\alpha$ predicted by \eqref{Benney_mu_critical} is indeed a conservative estimate of the necessary $\alpha$ required to stabilise the uniform film to perturbations of all wavelengths.

\subsection{Navier--Stokes equations}\label{subs:OS}

We can compute the linear stability of the Nusselt state in the two-dimensional Navier--Stokes equations, subject to distributed feedback controls, by a normal mode analysis. This analysis is well known in the absence of suction \citep{Floryan}. The addition of suction controls changes only one boundary condition in the resulting Orr-Sommerfeld system, and so only a brief description of the equations is presented here.

We perturb about the uniform state, writing 
\begin{equation}
 \begin{array}{cclcl}
    h &=& 1 &+& \epsilon \hat{H} \exp(ikx+\lambda t)
  \\
  u &=& \bar{u}(y) &+& \epsilon \hat{U}(y) \exp(ikx+\lambda t)
  \\
  v &=& 0&+ &\epsilon \hat{V}(y)\exp(ikx+\lambda t)
  \\
   p &=& \bar{p}(y) &+& \epsilon \hat{P}(y) \exp(ikx+\lambda t),
 \end{array}
\end{equation}
where $\bar{u}(y) = y(2-y)$ and $\bar{p}(y)$ correspond to the uniform film solution described in \eqref{uniform_film_solution},
and then linearise with respect to $\epsilon$.
The perturbation velocity components $\hat{U}(y)$ and $\hat{V}(y)$ can be expressed in terms of a streamfunction $\psi(y)$, so that
\begin{equation}
 \hat{U}(y) = -\psi'(y), \quad \hat{V}(y) = ik\psi(y),
\end{equation}
which immediately satisfies the mass conservation equation \eqref{NS_mass}.
The two components of the momentum equation \eqref{NS_xmom} and \eqref{NS_ymom} can then be combined to yield the Orr-Sommerfeld equation, which is a linear ordinary differential equation for $\psi$ in $0<y<1$:
\begin{equation}
\label{OS_bulk}
 \left( \frac{d^2}{dy^2} - k^2\right)^2 \psi = R[\lambda + i  k \bar{u}(y)]\left( \frac{d^2}{dy^2} - k^2\right) \psi - ik\bar{u}''(y)R \psi.
\end{equation}

The boundary conditions at the free surface are unaffected by $\alpha$, and after some manipulation involving \eqref{NS_xmom} to eliminate the fluid pressure, we obtain three boundary conditions at the free surface: 
\begin{equation}
-\psi'''(1)  + (R\lambda +ikR-3k^2) \psi'(1)  = 2ik\hat{H}\cot\theta + \frac{ik^3 \hat{H}}{C}, \quad
\psi''(1) = -2\hat{H}-k^2 \psi(1), \quad
 ik \psi(1) = (\lambda+ik)\hat{H}.
\end{equation}
The no-slip boundary condition on the wall yields
\begin{equation}
 \hat{U}(0) = -\psi'(0)=0
\end{equation}
and the responsive flux through the wall becomes the boundary condition
\begin{equation}
\label{OS_alpha}
 \hat{V}(0) = ik \psi(0) = -\alpha \hat{H}.
\end{equation}

When $k=0$, we can solve the system \eqref{OS_bulk}-\eqref{OS_alpha} for $\psi(y)$ and $\lambda$ analytically, and enumerate the eigenmodes. There is a single eigenmode that involves perturbations to the interface height (i.e. $\hat{H}\neq 0$), and for this eigenmode $\lambda = -\alpha$ at $k=0$. There are also an infinite number of shear eigenmodes which leave the interface position unperturbed. These eigenmodes are all stable, and the eigenvalue with the largest real part satisfies $\lambda R = -(\pi/2)^2$, irrespective of $\alpha$.

For $k\neq 0$, we solve the system \eqref{OS_bulk}-\eqref{OS_alpha} numerically. We can formulate the system as a generalised eigenvalue problem for $\psi$, $\hat{H}$ and $\lambda$, discretise the derivative operators on $\psi$ using finite differences or Chebyshev polynomials, and solve the resulting generalised eigenvalue problem using standard \textsc{Matlab} routines. Alternatively, we can formulate the complete system for the real and imaginary parts of $\psi$ and $\lambda$ as a boundary value problem in \textsc{Auto-07p}\citep{AUTO_07p} with $\lambda$ as a free parameter, as discussed by \citet{Kalliadasis}, though this is only useful for tracking a single eigenmode.

\begin{figure}
\includegraphics[width=0.8\textwidth]{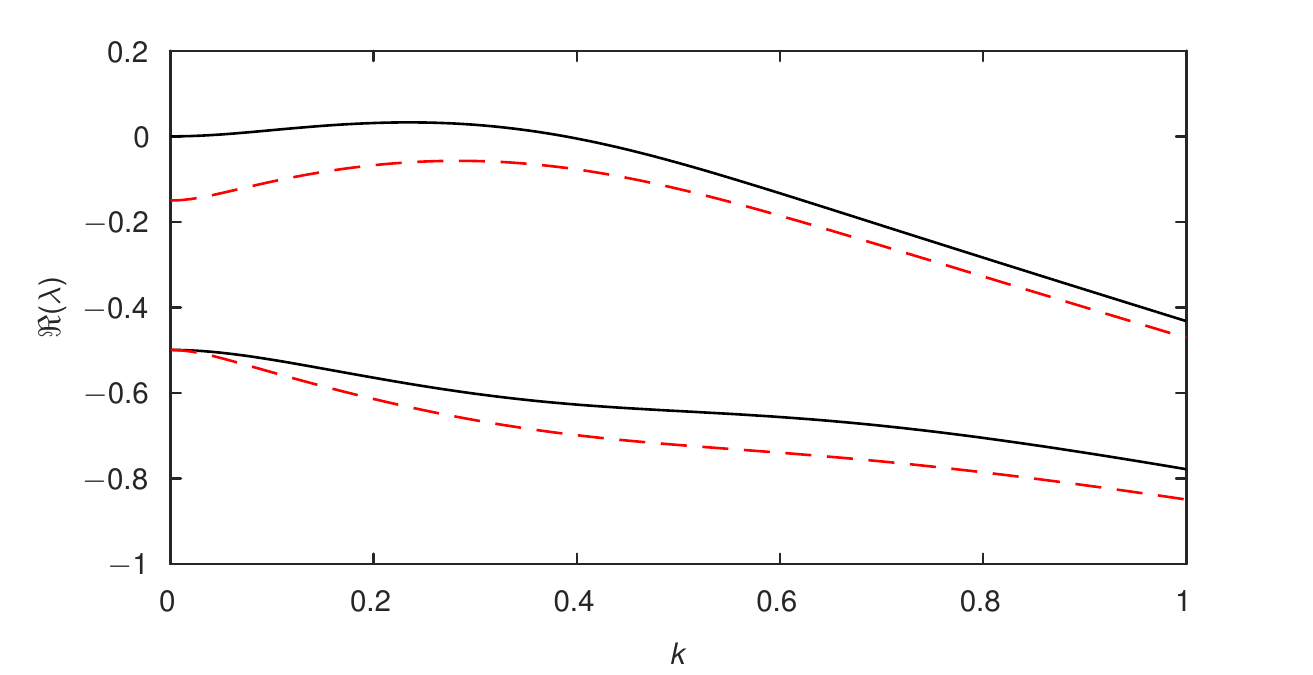}
 \caption{Results for the real part of the two Navier--Stokes eigenvalues $\lambda$ with largest real part, as a function of $k$, for $R=5$, $C=0.05$, $\theta=\pi/4$ and $\alpha=0$ (solid lines), and $\alpha=\alpha_B$ (dashed lines) from \eqref{Benney_mu_critical}. 
 \label{fig:OS_eigenvalues}}
\end{figure}

Results for $\lambda(k)$ are shown in Fig.~\ref{fig:OS_eigenvalues} for the two least stable eigenmodes. In the absence of controls, the Navier--Stokes results show a smaller cutoff wavenumber than the Benney and weighted residual results. As was the case for the weighted-residual equations, we find that introducing positive $\alpha$ decreases the real part of both eigenvalues shown, but has vanishing effect when $k=0$ on all but the least stable eigenmode. Furthermore, the critical $\alpha$ computed according to the Benney result \eqref{Benney_mu_critical} is again sufficient to stabilise the uniform state against perturbations of all wavelengths.

In Fig.~\ref{fig:Neutral_stability_map}, we show the critical $\alpha$ required so that $\Re(\lambda)\leq 0$ for all $k$ in the Navier--Stokes equations. This is computed in \textsc{Auto-07p}, with the condition that $\Re(\lambda)$ has both a turning point and a zero at the same value of $k$. When $\alpha<0.5$, this stability boundary is in good agreement with the weighted-residual results, with both predicting that the critical Reynolds number is increased substantially, from its uncontrolled value of 1.25 to around 50. Beyond this point, the weighted-residual results predict that the critical $R$ should continue to increase rapidly with $\alpha$. However, the Navier--Stokes results show a turning point in $R(\alpha)$, followed by a very slow decrease in $R$ as $\alpha$ is increased. 
This eventual deviation is not unexpected, given the wide range of Reynolds number spanned in this calculation.

\subsection{Initial value calculations}

\begin{figure}
 \includegraphics[width=0.8\textwidth]{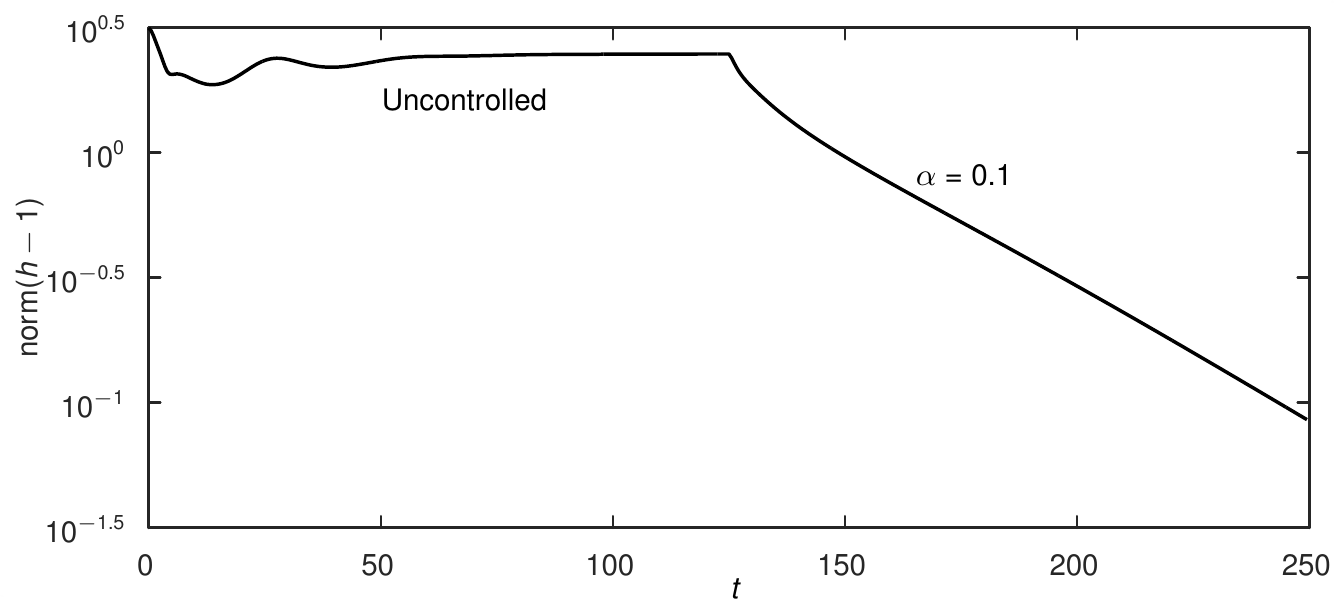}
 \caption{Results of an initial value calculation using the weighted residual equations, starting from a non-uniform, non-equilibrium state, which evolves without suction until $t=125$. For $t>125$, we enable feedback controls with $F = -0.1(h-1)$, and the system converges towards the uniform state.\label{fig:IVP_results}
 }
\end{figure}

Although we have demonstrated that the control parameter $\alpha$ can be chosen to make the uniform state linearly stable to perturbations of all wavelengths, it is not necessarily the case that the system will converge to the uniform state in nonlinear simulations. In Fig.~\ref{fig:IVP_results}, we show results of an initial value calculation of the weighted-residual system, starting from a finite-amplitude state that is neither a steady nor travelling-wave solution of the weighted-residual equations. We initially allow this state to evolve without controls, and find that the system moves towards a travelling wave state of finite amplitude. We then activate the feedback controls with $\alpha=0.1$, which is large enough that the uniform state is linearly stable. After the decay of transient behaviour, we observe that the distance of the solution to the desired state decays exponentially with respect to time, which is consistent with the expectation that the largest deviation is due to a single eigenmode which decays at constant rate. As the imposed injection is proportional to $h-1$, the control magnitude also decays exponentially with time, and would be expected to become vanishingly small at late times. However, although the amplitude of the applied injection and suction may become very small at late times, the feedback control scheme is still required to suppress the growth of small perturbations, and thus to ensure the linear stability of the system.

\subsection{Linear stability for phase-shifted distributed controls}
\label{sec:Phase_shift_analytical_eigenmodes}

\begin{figure}
 \includegraphics[width=0.8\textwidth]{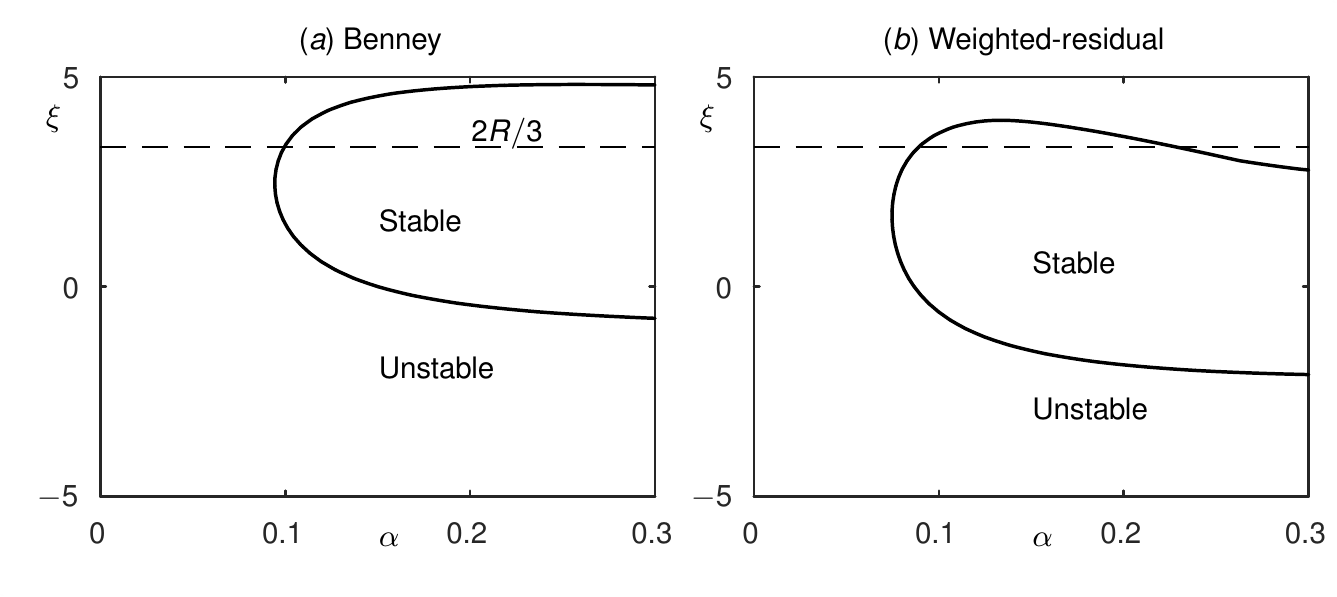}
 \caption{Linear stability properties of the uniform state as a function of the control strength $\alpha$ and the displacement $\xi$ between observer and actuator, with the control scheme \eqref{displaced_distributed_controls} for $R=5$, $\theta=\pi/4$, $C=0.05$. Stability results refer to perturbations of all wavelengths. The lowest $\alpha$ is required at a finite positive value of $\xi$. The dashed line shows the $O(k^2)$ optimiser in the Benney equations: $\xi =2 R/3$. \label{fig:Stability_map_alpha_phi}}
\end{figure}

As an initial step towards designing a more efficient system for feedback control, we can investigate the effect of shifting observations relative to actuators, still using a normal mode analysis.
We replace the control scheme~\eqref{uniform_control_scheme} with a scheme based on shifted observers:
\begin{equation}
\label{displaced_distributed_controls}
 F(x,t) = -\alpha [ h(x-\xi,t)-1].
\end{equation}
Here the real parameter $\xi$ is the distance between observer and actuator. 
Positive $\xi$ means that the observers are displaced upstream relative to the position at which the injection is applied. This scheme introduces no favoured $x$ locations, and so the eigenmodes can still be written as
\begin{equation}
 h = 1+ \epsilon \hat{h} \exp(ikx+\lambda t) + O(\epsilon^2), \quad q = 2/3 + \epsilon \hat{q} \exp(ikx+\lambda t).
\end{equation}
We then find
\begin{equation}
 F = -\alpha e^{-ik\xi} \epsilon \hat{h} \exp(ikx+\lambda t).
\end{equation}
We thus simply replace $\alpha$ by $\alpha \exp(-ik\xi)$ in \eqref{Benney_dispersion} and \eqref{WR_lambda_eq} to understand the effect of $\xi$ on the eigenvalues of the Benney and weighted-residual models respectively. For both models, we can perform a numerical search to calculate the boundary of the region in $\alpha$-$\xi$ space where the uniform state is stable to perturbations of arbitrary wavelengths, as shown in Fig.~\ref{fig:Stability_map_alpha_phi}; for the parameters in this figure, we find that choosing $\xi\approx 2$ has the best stabilising effect in both models, in the sense that a stable uniform state is obtained at the lowest value of $\alpha$.
There are some differences between the results for the two models: the effect of positive $\xi$ is less pronounced in the weighted-residual model than in the Benney calculations, and in fact for the weighted-residual model, choosing positive $\xi$ eventually becomes less stabilising as $\alpha$ is increased.

In order to understand why the control scheme is most effective when actuators are displaced upstream by a finite distance, 
we can expand the Benney eigenvalue under the assumption that $k\xi$ is small, to reach
\begin{equation}
 \Re(\lambda) = -\alpha \left(1 + k^2\xi\left[ \frac{2R}{3} -  \frac{\xi}{2} \right]\right) + \frac{8k^2}{15}\left(R-\frac{5\cot\theta}{4} - \frac{5k^2}{8C}\right).
\end{equation}
To maximise the effect of $\alpha$, we should choose $\xi = 2R/3$, which provides a reasonable estimate of the optimal $\xi$, as shown in Fig.~\ref{fig:Stability_map_alpha_phi}. This should become a better estimate as $R\rightarrow R_0$, so that the unstable $k$ move towards zero.

\section{Applying controls via point actuators}
\label{sec:Actuators_and_observations}

A physically important question that we wish to address next is the application of suction controls using point actuators, and based on a limited number of observations of the system state.
Here we consider only behaviour within a spatial period of length $L$, and only stabilisation of the uniform state.

We are given the localised actuator functions $\Psi_m(x)$, so that
\begin{equation}
 F(x,t) = \sum_{m=1}^M b_m(t) \Psi_m(x),
\end{equation}
where the $M$ coefficients $b_m(t)$ are to be determined from $P$ discrete observations $y_p(t)$ of the interface height:
\begin{equation}
 y_p(t) = \int_0^L \Phi_p(x) (h(x,t)-1) \,\mathrm{d}x.
\end{equation}

We note that the explicit $x$-dependence of the system that arises from localised actuators and observers means that the system is no longer translationally invariant in $x$, and so linear stability properties of even a uniform film in the Navier--Stokes equations cannot be obtained by a normal mode analysis. Instead, we derive most of our control strategies using the Benney model, and use the weighted-residual model as a black box experiment to represent the additional complexities of the full physical system subject to controls derived using a low order model.

As a starting point, we suppose that the controls are a linear function of the instantaneous observation, which is known as a static observation scheme. In the most general form, we can then write 
\begin{equation}
\label{Point_actuator_F}
 F = \Psi K \Phi(h-1).
\end{equation}
Here the operator $\Phi$
describes observations of the system, $\Psi$ represents the shape of the actuators, and $K$ is the control operator which we are free to choose based on our knowledge of $\Phi$, $\Psi$ and the system dynamics.  We will use $M$ linearly independent actuators and $P$ observations, which are the ranks of $\Psi$ and $\Phi$ respectively.
In a discretised form, $\Psi$ and $\Phi$ are matrices of size $N\times M$ and $P\times N$ respectively. The matrix $K$ has size $M\times P$, and we may choose all of its entries.
Given this form for $F$,  we can compute the linear stability of a given steady state by replacing $\alpha$ with $-\Psi K \Phi$ in the eigenvalue problems described in Sec.~\ref{sec:Eigenvalue_problems}.

\subsection{Choice of point actuators and observers}

We choose to use $M$ equally-spaced actuators, which are each periodic with period $L$ and locally behave as Dirac $\delta$-functions, so that
\begin{equation}
\label{Psi_delta}
 \Psi_m(x) = \delta(x-x_m), \quad x_m = mL/M.
\end{equation}
We similarly use $P$ equally-spaced observer functions, which are displaced upstream by a distance $\xi$ from the actuator positions, so that
\begin{equation}
\label{Phi_delta}
 \Phi_p(x) = \delta(x-x_p), \quad x_p = pL/P-\xi.
\end{equation}

\begin{figure}
 \includegraphics[width=0.8\textwidth]{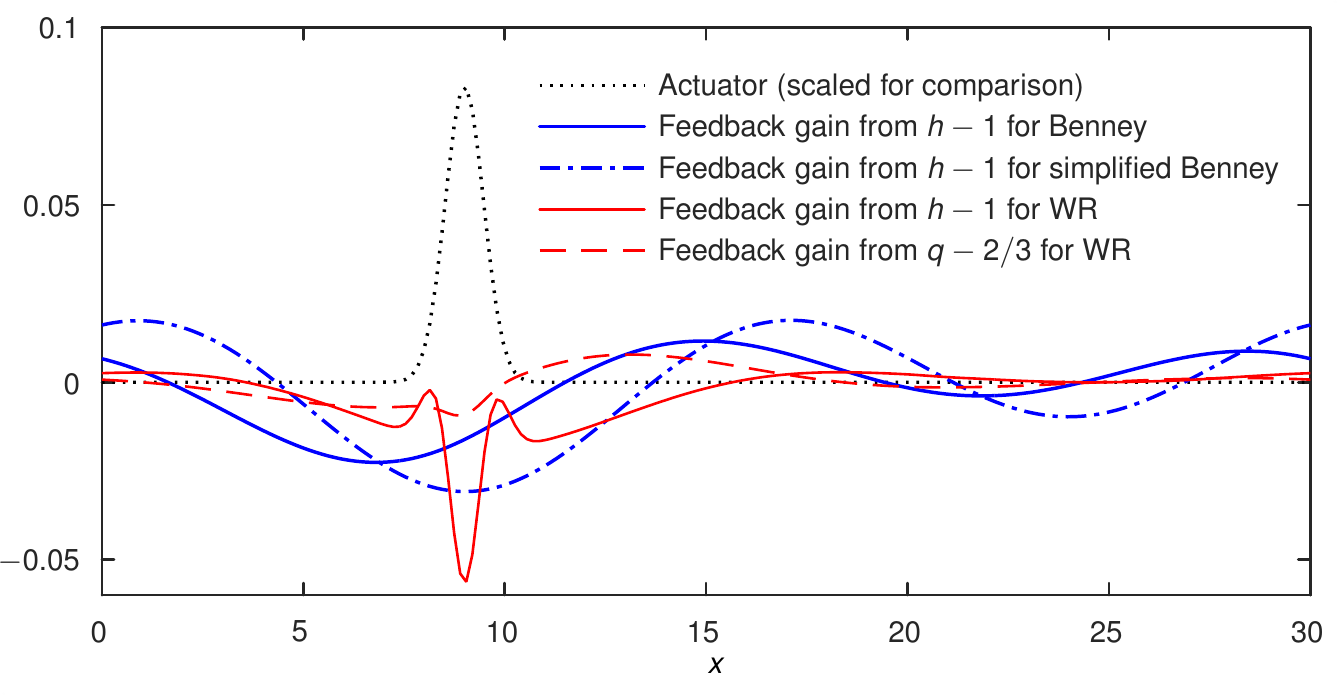}
 \caption{A typical row of the matrix $K$, or feedback gain, obtained by the LQR algorithm, with 5 equally spaced actuators, with shape smoothed according to \eqref{smoothed_shape} with $w=0.1$, and shown by the dotted line here.  The cost parameter is $\mu=0.1$, and for the weighted-residual equation, the same cost weighting is associated with $q-2/3$ as for $h-1$.
 \label{fig:compare_LQR_WR_Benney}
 }
\end{figure}

For our numerical calculations, we replace $\delta(x)$ in \eqref{Psi_delta} and \eqref{Phi_delta} by the smoothed, periodic function $d(x)$, defined by
\begin{equation}
\label{smoothed_shape}
 d(x) = \exp\left[ \frac{\cos(2\pi x)-1}{w^2}\right].
\end{equation}
One such actuator shape function is plotted in Fig.~\ref{fig:compare_LQR_WR_Benney} for $w=0.1$.
We normalise the smoothed functions $d(x)$ so that each actuator and observer shape function has integral 1 over the interval $[0,L]$, and so $d(x)\rightarrow \delta(x)$ as $w\rightarrow 0$.

\subsection{Proportional control}

If the number of actuators is equal to the number of observers, one of the simplest methods to choose the suction/injection profile is to link each actuator to a neighbouring observer, setting 
\begin{equation}
\label{Single_links}
 b_m(t) = -\alpha y_m(t)
\end{equation}
where the positive control amplitude $\alpha$ acts in a similar way to the control parameter $\alpha$ in Sec.~\ref{sec:Control_to_uniform}. In terms of the generalised eigenvalue problems, we simply set $K = -\alpha I$.
If all actuators, and all observers, are equally spaced, the control scheme is specified entirely by $\alpha$ and the displacement $\xi$ between actuator and observer.
In Sec.~\ref{sec:Control_to_uniform}, we considered the continuous analogue of this scheme, with feedback at every point proportional to the interface height at that point only. We found that positive $\alpha$ had a stabilising effect on the system dynamics according to both long-wave models and also in the Navier--Stokes equations. 

\begin{figure}
 \includegraphics[width=0.8\textwidth]{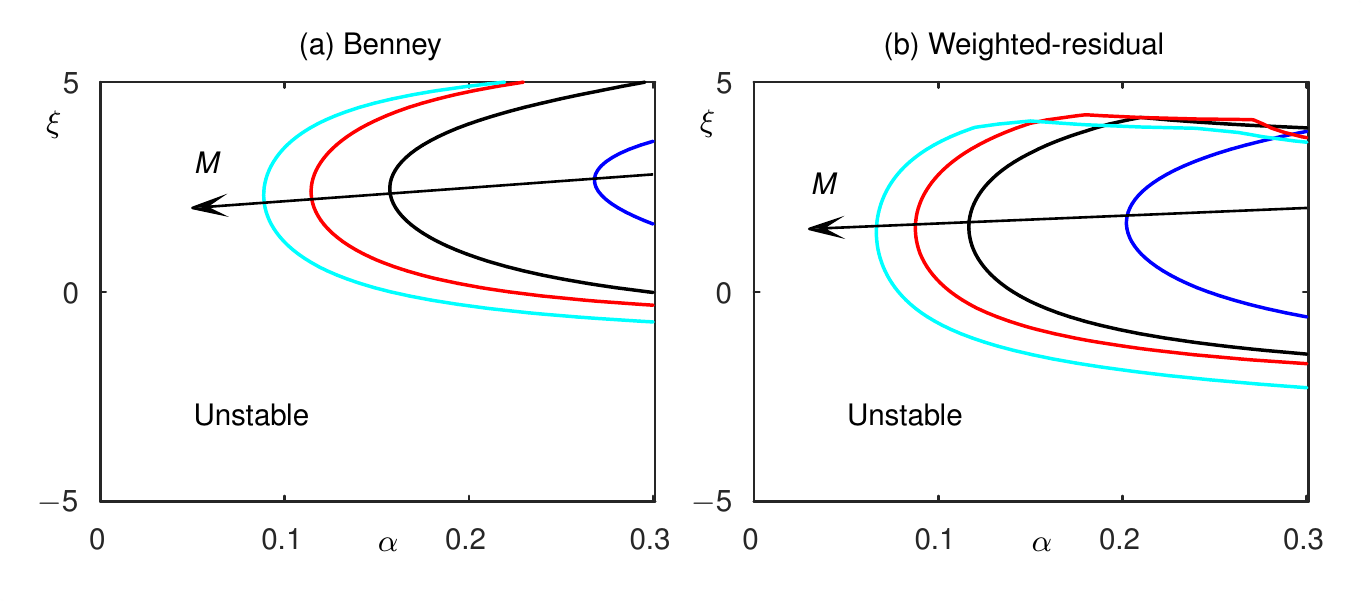}
 \caption{Stability results as the control amplitude $\alpha$ is varied, with a phase shift $\xi$ between actuator and observer. There are $M$ equally-spaced actuators, and $P=M$ equally-spaced observers, each smoothed according to \eqref{smoothed_shape} with $w=0.1$, and results are shown for $M=P=3,5,7,9$. The largest stable region occurs for $M=9$.
 As is the case for distributed actuators (see appendix \ref{sec:Phase_shift_analytical_eigenmodes} and Fig.~\ref{fig:Stability_map_alpha_phi}), the best stabilisation occurs at a moderate, positive value of $\xi$, so that the observers are positioned upstream relative to the actuators. 
\label{fig:Phase_shift_stability_5_actuators} }
\end{figure}

When applying the proportional control scheme \eqref{Single_links} with localised observers and actuators, the eigenmodes are not sinusoidal, and so we calculate the linear stability properties numerically by solving an eigenproblem; note that this calculation only allows for perturbations with wavelength at most $L$. We find that increasing $\alpha$ has a stabilising effect on the uniform film, and that the value of $\alpha$ required to obtain a linearly stable state decreases when increasing the number of actuators $M$ and observers $P$ (see Fig.~\ref{fig:Phase_shift_stability_5_actuators}).

We can also investigate the effect of the displacement $\xi$ between the observers and actuators on the linear stability of the uniform state. 
As discussed in Sec.~\ref{sec:Phase_shift_analytical_eigenmodes}, the uniform state is most easily stabilised by distributed controls when $\xi\approx 2$, and when $R$ is close to $R_0$, this best choice for $\xi$ is given by $\xi \sim 2R/3$.
However, the use of localised observers and actuators introduces a natural lengthscale $L/M$, which is the distance between neighbouring observers, and for the analysis in this section, we also have the lengthscale $L$ of the imposed periodicity.  
We can numerically calculate the effect of the displacement $\zeta$ on the linear stability of the uniform film, subject to the control scheme \eqref{Single_links}, by solving an eigenvalue problem for each $\xi$ and $\alpha$. Fig.~\ref{fig:Phase_shift_stability_5_actuators} shows the stability boundaries in $\alpha$-$\xi$ space for $M=P=3,5,7,9$. We find that a stable state can be obtained at the smallest $\alpha$ when $\xi\approx 2$, which is comparable to the results of the calculations for distributed controls and observations shown in Fig.~\ref{fig:Stability_map_alpha_phi}, despite the additional lengthscales present in the system with localised observers and controls. The magnitude of $\alpha$ required to stabilise the uniform state generally decreases as $M=P$ is increased, but even for just three actuators, we can stabilise the uniform film state by choosing a sufficiently large $\alpha$ with $\xi \approx 3$.

\subsection{Linear-quadratic regulator with full observations}
\label{sec:LQR}

The control scheme described in the previous subsection only allows each actuator to communicate with a single observer. We should be able to obtain better control by allowing data from all observers to be combined before determining the actuator amplitudes; we will still consider linear control, but allow all entries of the $M\times P$ matrix $K$ to be non-zero. This more general scheme can also encompass situations where $M\neq P$.

The statement that the full system state can be observed is a stringent constraint; for the weighted-residual model this requires simultaneous information regarding $h(x,t)$ and $q(x,t)$, and in the Navier--Stokes system, the full system state includes two components of the velocity field along with the interface height.
Notwithstanding the difficulties of obtaining full observations, if we are somehow able to observe the full system state, a variety of algorithms from control theory can be used to compute the controls. Here we choose to use the linear-quadratic regulator (LQR) algorithm \citep{Zabczyk1992}, which determines $K$ so as to minimise a cost functional associated with control amplitudes and the deviation of the system from the flat state.

The LQR algorithm  is designed for the system
\begin{equation}
\label{LQR_system}
 \frac{dz}{dt} =  Jz + \Psi u, \quad u = Kz,
\end{equation}
where $z$ and $u$ are vectors, the matrices $J$ and $\Psi$ are given, and we wish to choose the matrix $K$ 
in order to minimise the cost $\kappa$ defined by
\begin{equation}\label{eq:cost}
 \kappa = \int_0^\infty \left(z^T U z + u^T V u\right)  \, \mathrm{d}t,
\end{equation}
where $U$ and $V$ are given symmetric, positive definite matrices that define the relative cost associated with different solution components. A minimiser $K$ of the cost \eqref{eq:cost} subject to the system \eqref{LQR_system} is strongly connected to a solution, if it exists, of an algebraic Ricatti equation:
\begin{equation}\label{eq:Ricatti}
U + PJ + J^TP - P\Psi V^{-1}\Psi^TP = 0,
\end{equation}
in which the unknown $P$ is a nonnegative definite matrix.  If $\tilde{P}$ is a solution to \eqref{eq:Ricatti} and $\tilde{P}- P$ is negative definite for all other solutions $P$, then $\tilde{P}$ is called a minimal solution to \eqref{eq:Ricatti} and $K = -V^{-1}\Psi^T\tilde{P}$ minimises the cost functional \eqref{eq:cost}.
Furthermore, in \citet{Zabczyk1992}, it is proved that if the pair $(J,\Psi)$ is controllable and $U = C^TC$, where the pair $(J,C)$ is observable (see Appendix~\ref{appB} for definitions of controllability and observability) then the algebraic Ricatti equation \eqref{eq:Ricatti} has exactly one solution $P$, and the matrix $J-\Psi V^{-1}\Psi^TP$ is stable.

For simplicity, we use the following cost functional, in terms of our variables, 
\begin{equation}
\label{kappa_definition}
 \kappa = \int_0^\infty \int_0^L \left\lbrace\mu (h-1)^2 + (1-\mu) F^2\right\rbrace \, \mathrm{d}x \, \mathrm{d}t. 
\end{equation}
For a given physical system, the control scheme is a function of the single parameter $\mu \in (0,1)$. The choice of $K$ and the resulting system eigenvalues are dependent on $\mu$, but a stable system should be obtained for any $0<\mu<1$. Row $m$ of the matrix $K$ determines the amplitude of actuator $m$:
\begin{equation}
 b_m(t) = \sum_{n=1}^{N} K_{mn} (h_n(t)-1).
\end{equation}

Fig.~\ref{fig:compare_LQR_WR_Benney} shows one such row, or feedback gain, computed using the LQR algorithm, as implemented  using the \textsc{Matlab} LQR function, for the Benney and weighted-residual equations.
The LQR algorithm gives very smooth control input functions for the Benney equation. The largest part of the input function is localised slightly upstream of the actuator location when using the full Benney equation \eqref{eq:Benney}, or more centrally when using the simplified version \eqref{eq:Benney_simplified}. We can insert the Benney controls directly into the weighted-residual model, and in fact still obtain a stable state.

We can also use the LQR algorithm to calculate controls for the weighted-residual model, but the controls require observations of both $h$ and $q$. We also note that the control input functions (Fig.~\ref{fig:compare_LQR_WR_Benney}) have relatively sharp edges near the width of the actuator. 
The full LQR controls are able to stabilise the uniform state in the weighted-residual model, and for our test case the maximum real part of any eigenvalue is $-5.62 \times 10^{-2}$. 
Realistically, we are unlikely to have access to observations of both $h$ and $q$, and so it would be desirable to approximate $q$ from our observations of $h$ using a low order model.
The simplest method is to suppose that $q=2/3$, in effect discarding the control component from $q$. We find that this yields a linearly stable system, but the maximum real part of any eigenvalue is then $-5.09\times 10^{-3}$, so that convergence towards the uniform state would be very slow. We can recover the information regarding the $q$ controls by supposing that $q = 2h^3/3$ (which is the leading order term in the long-wave flux \eqref{eq:Benney}), and so $\hat{q} = 2\hat{h}$. The largest growth rate is then $-5.64\times 10^{-2}$, which is comparable to the growth rate obtained when the flux $q$ can be fully observed.

\subsection{Dynamical observers for a finite number of observations}
\label{sec:Evolving_controls}

For the LQR methodology described above, full observations of the system state are assumed to be available.
The system is specified by the interface shape in the Benney equation, but in the weighted-residual equations we also require full knowledge of the total down-stream flux at each streamwise location. Furthermore, for the Navier--Stokes equation we would need to know the instantaneous velocity at every point within the fluid. Such knowledge is unrealistic, and so we now consider the case where the only system observations available are those of the interface height, $h$, at only a finite number of points within the periodic domain. 
In the previous subsection, we showed that if full observations are available, standard algorithms, such as LQR, can be used to construct a control matrix $K$ for the instantaneous control scheme \eqref{Point_actuator_F} so  that localised actuators can be used to stabilise the uniform state. Alternatively, if distributed actuators can be applied, the LQR algorithm can also be used to calculate a control scheme subject to localised observers (see discussion in Appendix). However, if there are restrictions on both actuators and observers, it is not always possible to construct a control scheme based on \eqref{Point_actuator_F} so that the uniform state is linearly stable. Instead, we turn to a system of dynamical observers, in which both current and historical observations are used to determine the controls.

The principle of the approach described here is to construct an approximation of the system state which is continually corrected based on the observations available. We focus our effort on approximating the coefficients of those modes which are unstable in the uncontrolled system. We use the dynamic method described by \citet{Zabczyk1992} and applied for the KS equation in \citet{Armaou2000}, where the predictions evolve in time according to our understanding of the linearised system behaviour in the form of its Jacobian matrix and the system amplitudes, and the predicted amplitudes are corrected according to our observations. This is in contrast to a static observation scheme \eqref{Point_actuator_F}, where the controls are calculated only from the most recent set of observations.

After transformation to Fourier space, we can describe the evolution of a small perturbation $\tilde{h}$ in the (simplified) Benney equation \eqref{eq:Benney_simplified} by
\begin{equation}
 \frac{d\tilde{h}}{dt} = \tilde{J}\tilde{h} + \tilde{F}.
\end{equation}
In the absence of suction, the system has no preferred positions, and so the eigenvectors of $J$ are Fourier modes, and the transformed Jacobian matrix $\tilde{J}$ is diagonal. We reorder the wavenumbers so that the unstable eigenmodes of $J$ appear first:
\begin{equation}
\frac{d\tilde{h}}{dt} = \tilde{J}\tilde{h} + \tilde{F} = 
 \begin{pmatrix}
  \tilde{J}_u & 0 \\
  0 & \tilde{J}_s
 \end{pmatrix}
\tilde{h} + \tilde{F},
\end{equation}
where the subscripts $u$ and $s$ correspond to unstable and stable modes, respectively.
We wish to control to the state $\tilde{h}=0$.

To stabilise the zero state of this system, we would ideally leave the stable modes untouched, while choosing $F$ to react to the unstable modes.
This can be achieved by letting
\begin{equation}
\tilde{F} = \tilde{\Psi}\tilde{K}\tilde{h} = 
\begin{pmatrix}
\tilde{\Psi}_u \\ \tilde{\Psi}_s \end{pmatrix}\tilde{K}\tilde{h},
\end{equation}
so that
\begin{equation}
\frac{d}{dt}\begin{pmatrix}
            \tilde{h}_u \\ \tilde{h}_s 
            \end{pmatrix}
=
 \begin{pmatrix}
  \tilde{J}_u & 0 \\
  0 & \tilde{J}_s
 \end{pmatrix}
\begin{pmatrix}
             \tilde{h}_u \\ \tilde{h}_s 
            \end{pmatrix}
 +
 \begin{pmatrix}
\tilde{\Psi}_u \tilde{K} & 0 \\ \tilde{\Psi}_s \tilde{K} & 0
 \end{pmatrix}
\begin{pmatrix}
      \tilde{h}_u \\ \tilde{h}_s 
    \end{pmatrix} 
    =  \begin{pmatrix}
  \tilde{J}_u + \tilde{\Psi}_u \tilde{K} & 0 \\
  \tilde{\Psi}_s \tilde{K} & \tilde{J}_s
 \end{pmatrix}
 \begin{pmatrix}
      \tilde{h}_u \\ \tilde{h}_s 
    \end{pmatrix}. 
\end{equation}
The matrix on the right-hand side of the eigenvalue problem is lower triangular by blocks, and the block $\tilde{J}_s$ is diagonal. The eigenvalues and eigenvectors corresponding to $\tilde{J}_s$ are thus unchanged by $\tilde{F}$.

The remaining task is to stabilise the subsystem 
\begin{equation}\label{eq:unstable_subsystem}
\frac{d\tilde{h}_u}{dt} =\tilde{J}_u\tilde{h_u} + \tilde{\Psi}_u\tilde{K}_u\tilde{h}_u.
\end{equation}
To choose the matrix $\tilde{K}_u$, we use the LQR algorithm on the subsystem \eqref{eq:unstable_subsystem}, which has size equal to the number of unstable modes, $M$. However, to apply these controls, we need to approximate $z=\tilde{h}_u$ based on our observations.
We can write our discrete set of observations as $y= \Phi(h-1)$, $\tilde{y} = \tilde{\Phi}\tilde{h} = \tilde{\Phi}_u\tilde{h}_u + \tilde{\Phi}_s\tilde{h}_s$.

We can obtain a good approximation of $z$ by considering a set of ordinary differential equations:
\begin{equation}
\label{eq:z_evolve_0}
\frac{dz}{dt} = (\tilde{J}_u  + \tilde{\Psi}_u \tilde{K}_u) z + L(y-\bar{y}) = \left(\tilde{J}_u + \tilde{\Psi}_u\tilde{K}_u - L\tilde{\Phi}_u\right) z  + Ly, \quad \bar{y} = \Phi_u z.
\end{equation}
Here $\bar{y}$ is the expected set of observations based on our current approximation to the system, and the 
 $L(y-\bar{y})$ term indicates a correction based on our actual observations. Once we know $z$, we can set $\tilde{F} = \tilde{\Psi} \tilde{K}_u z$.  
However, we still need to choose the matrix $L$ in order that $z$ will converge rapidly to $\tilde{h}_u$. We define an error term: $\tilde{e} = \tilde{h}_u - z$, and after several substitutions we find that $\tilde{e}$ is governed by
\begin{equation}
\label{error_equation}
 \frac{d\tilde{e}}{dt} = (\tilde{J}_u - L\tilde{\Phi}_u) \tilde{e} - L \Phi_s \tilde{h}_ss = Y \tilde{e} - L \Phi_s \tilde{h}_s.
\end{equation}
To obtain rapid convergence of our estimator $z$ towards the true system state, we need the eigenvalues of the matrix $Y$ to have large and negative real part, and we can obtain a suitable matrix $L$ by using the LQR algorithm. If these conditions on the eigenvalues of $Y$ are satisfied, it can be proved that the solution $z$ to equation \eqref{eq:z_evolve_0} converges exponentially to the true coefficients $\tilde{h}_u$ of $h$ as long as the initial guess is sufficiently good.
 Furthermore, if the real part of these eigenvalues is sufficiently large (in absolute value), then we can write $Y = \widetilde{Y}/\epsilon$ for small $\epsilon$ and, by multiplying \eqref{error_equation} by $\epsilon$, can obtain a system of equations in the standard singularly perturbed form \citep{Kokotovic1986}. This system possesses an exponentially stable fast subsystem (the equation for $z$) and an exponentially stable slow subsystem (the (stabilised) linearised equation for $h$), which implies that the system \eqref{eq:stabilised_with_observations} is exponentially stable.

We can rewrite the complete system in real space, to determine the behaviour of the nonlinear initial value problem:
\begin{subequations}\label{eq:stabilised_with_observations}
\begin{align}
\label{mass} h_t + q_x &= F(x,t), \\
\label{qBenney} q(x,t) &=  \dfrac{h^3}{3}\left( 2 -2 h_x\cot\theta + \dfrac{h_{xxx}}{C}\right)+ \dfrac{8 R h^6h_x }{15},\\
\label{Control} F(x,t) &= \mathcal{F}^{-1}\tilde{\Psi}\tilde{K}_u z,\\
\label{z_evolution} \dfrac{dz}{dt} &= \left(\tilde{J}_u + \tilde{\Psi}_u\tilde{K}_u - L\tilde{\Phi}_u\right) z  + Ly,\\
\label{y_observ} y & =  \Phi(h-1),
\end{align}	
\end{subequations}
where $\mathcal{F}$ is the Fourier transform operator. It can be seen from equations~\eqref{Control}-\eqref{y_observ} that the feedback control $F$ is calculated only from those observations of the system state attainable through the matrix $\Phi$.

\begin{figure}
\includegraphics[width=0.8\textwidth]{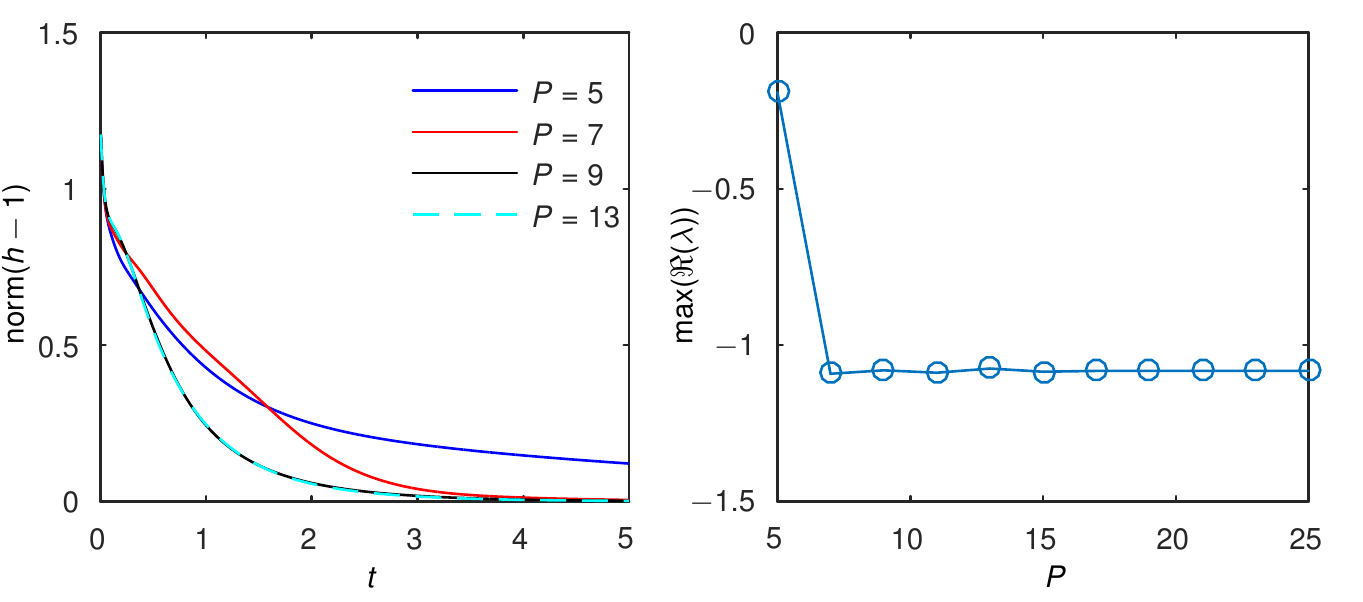}
 \caption{Distance between current solution and uniform film state as a function of time for $M=5$ actuators, with $P$ observers; and maximum real part of the eigenvalues of the system \eqref{eq:stabilised_with_observations} as a function of $P$. The actuator and observer shapes are as described by \eqref{smoothed_shape}, with $w=0.1$ and $\xi=0$. The initial condition is $h=1 + 0.3\cos(2\pi x/L) + 0.1\sin(4\pi x/L)$, with $L=30$. \label{fig:IVP_evals_evolving}}
\end{figure}

\begin{figure}
\includegraphics[width=0.8\textwidth]{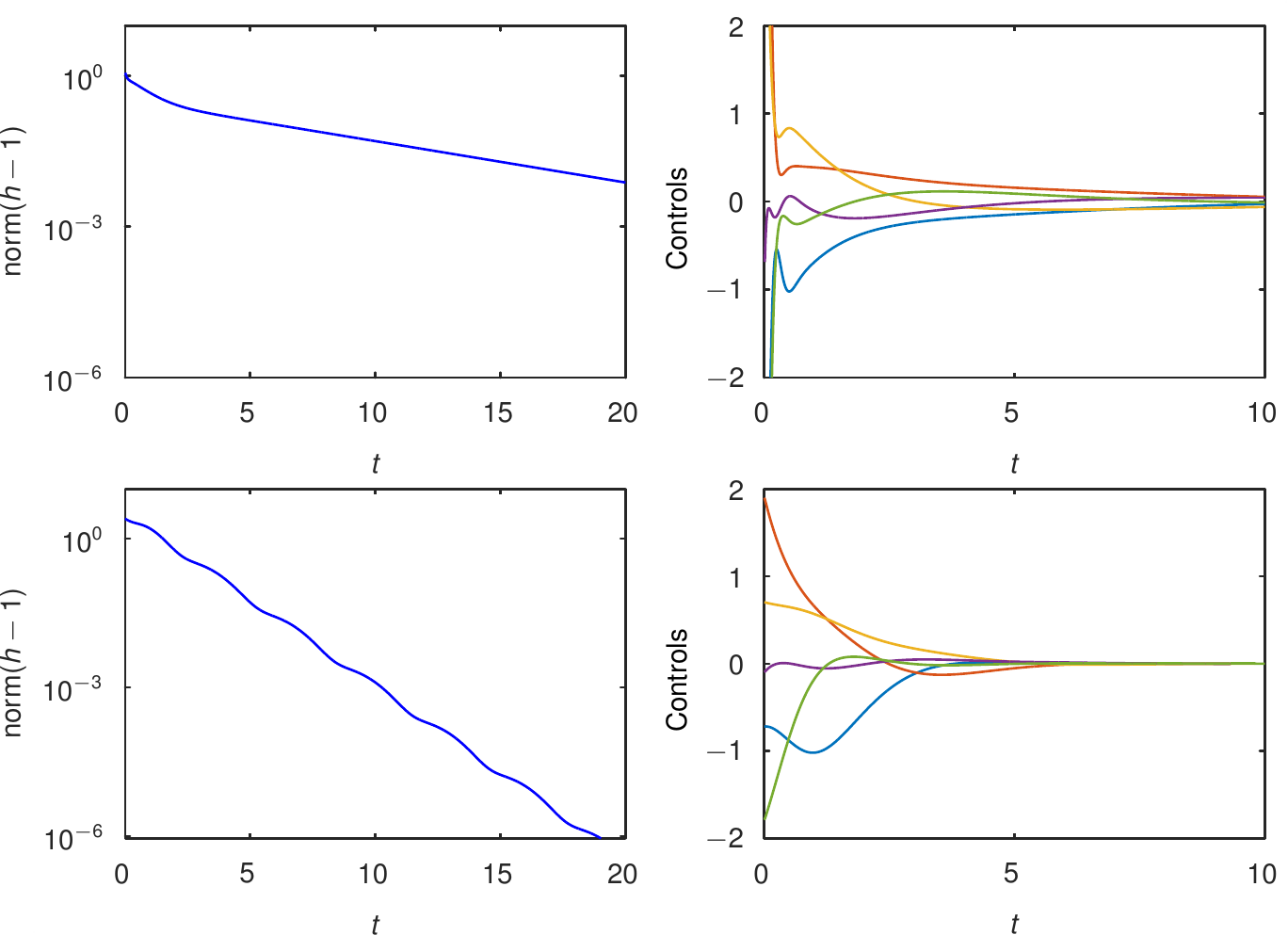}
 \caption{Semi-log plot of the distance between the current and flat states (left) and amplitudes of controls as a function of time (right), for $M=5$. For the upper row of figures, we use $P=5$ observations, while for the lower row, we use full knowledge of the interface height $h$. \label{fig:IVP_semilog_and_controls}}
\end{figure}

It is necessary to alter, and hence approximate, all of the unstable eigenmodes of the system in order to stabilise the uniform state, and so the size of $z$ must be equal to or greater than the number of unstable modes. We expect to achieve better performance as the number of tracked and stabilised modes is increased. The number of actuators $M$ need not be equal to the number of observers $P$, and Fig.~\ref{fig:IVP_evals_evolving} shows system eigenvalues as $P$ is increased for $M=5$ (note that $P$ is odd). We find that choosing $P=7$ gives much faster convergence than $P=5$, but further increases in $P$ have negligible effect on the eigenvalues. However, nonlinear initial value simulations of the system \eqref{eq:stabilised_with_observations} benefit from taking $P=9$.
In Fig.~\ref{fig:IVP_semilog_and_controls}, we compare nonlinear initial value calculations for $M=5$, based on $P=5$ and on full observations. We find that much faster convergence is obtained with full observations.

The system \eqref{eq:stabilised_with_observations} is presented for the simplified Benney equation \eqref{eq:Benney_simplified}. The analysis can be extended to include cross flow effects present in \eqref{eq:Benney} by left-multiplying $\Psi$ by $(I-\partial_x Z_F)$ before computing $\tilde{\Psi}_u$.
The Benney control scheme can be implemented in the weighted-residual equation by simply replacing equation~\eqref{qBenney} by \eqref{eq:WR}, but we cannot be certain that the resulting system will be linearly stable. For our test case, we find that even the linear stability of the uniform state in the weighted-residual equations is sensitive to $P$, with $P=5$ stable, but $P=7$ unstable. A full analysis of the approximately-controlled weighted-residual equation is a topic for future work.

\section{Controlling to non-uniform solutions with distributed controls}
\label{sec:Non_uniform}

Feedback controls of the form $F = -\alpha(h-H)$ can also be used to drive the system towards non-uniform states, by setting the target state $H$ to be spatially varying. We would like to know whether the state $h=H$ is always reached, and whether this state is stable. As $F=0$ when $h=H$, the system can only remain in this state if $h=H$ is an exact solution of the equations in the absence of suction. Small perturbations about $H$ are always affected by the feedback controls, and so $\alpha$ will change the linear stability properties of the state $H$. We will discuss the system dynamics when $H$ is not an exact solution of the governing equations in Sec.~\ref{sec:Imperfect_model_knowledge}.
The extension of the localised actuator control scheme developed in the previous section to stabilise a non-uniform state is a non-trivial task, as discussed in Sec.~\ref{sec:Nonuniform_actuators}, and is left for future work.

\subsection{Travelling waves}
\label{sec:Travelling_waves}

The long-wave systems support non-uniform travelling wave solutions, of the form $h = H(x-Ut)$, where $U$ is the propagation speed. Travelling waves undergo bifurcations (see e.g. \citet{Oron_Gottlieb_Benney_bifurcations}), and may be stable or unstable in the corresponding moving frame.
It is important to note that the shapes and bifurcation structure of travelling waves differ between the models. 
If the target state $H$ is an exact travelling wave solution to the equations in the absence of suction, then the state $H$ is also a travelling wave solution to the same equations with $F =- \alpha[h-H(\zeta)]$, and thus the application of controls affects the stability but not the shape or speed of the targeted travelling wave.

\begin{figure}
 \includegraphics[width=0.8\textwidth]{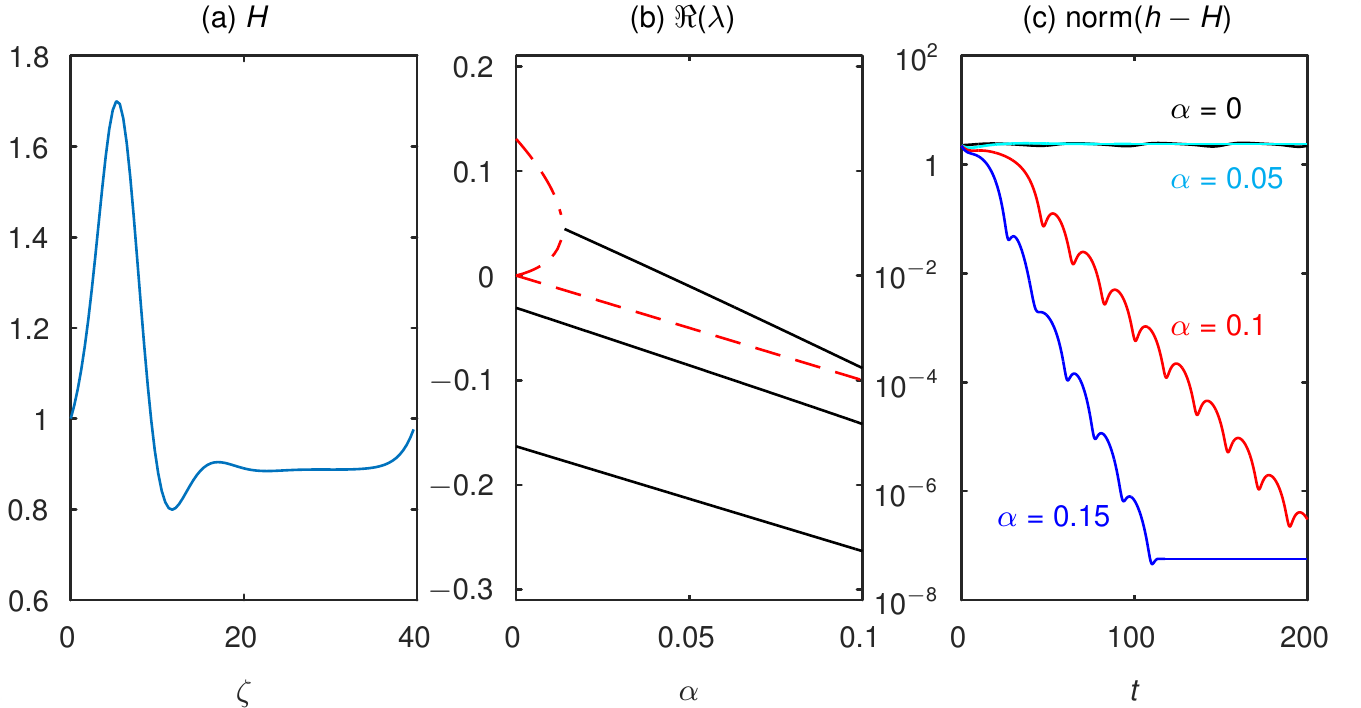}
 \caption{(a) A travelling wave solution to the Benney equation, for $R=2$, $\theta=\pi/4$, $C=0.05$, $U=2.82$. (b) The real part of the seven complex eigenvalues with largest real part, as $\alpha$ is increased. Real eigenvalues are shown by red dashed lines, while black solid lines indicate the real part of complex conjugate pairs. Neutral stability occurs at $\alpha = 0.0434$. (c) Results from nonlinear initial value calculations, starting from close to a uniform film, controlling towards the solution shown in (a), for $\alpha=0$, $\alpha = 0.05$, $\alpha = 0.1$, $\alpha=0.15$. Convergence to $H$ is only achieved in the two latter cases. \label{fig:Travelling_wave_calculations}}
\end{figure}

Fig.~\ref{fig:Travelling_wave_calculations}(a) shows an unstable travelling wave solution to the Benney equation. For simplicity, we limit perturbations to those periodic with the same spatial period as the travelling wave. 
In order to compute the stability of travelling waves, we transform to the frame moving at speed $U$, and then identify $x$ with $\zeta$. For the Benney equation, the generalised eigenvalue problem \eqref{eq:small_e_value} becomes
\begin{equation}
  \lambda \hat{h} = \left\lbrace U \partial_{\zeta}  -\partial_\zeta Z_h - [I-\partial_\zeta Z_F]\alpha \right\rbrace\hat{h}.
\end{equation}
We note that if $Z_F=0$, which is the case for the simplified Benney equation \eqref{eq:Benney_simplified}, then the eigenvalues $\lambda$ are shifted by $-\alpha$, and the eigenvectors of the system are unchanged from those in the absence of controls.
However, if we are using the standard Benney equation \eqref{eq:Benney} or the weighted-residual system, the effect of $\alpha$ on the eigenvalues is more complicated than a simple shift, and we solve the eigenvalue problem numerically to determine the effect of $\alpha$ on the linear stability properties of the non-uniform travelling waves.

Fig.~\ref{fig:Travelling_wave_calculations}(b) shows the real part of the seven most unstable eigenmodes as a function of $\alpha$ when considering the linear stability of the travelling wave shown in Fig.~\ref{fig:Travelling_wave_calculations}(a). When $\alpha=0$, this travelling wave is unstable, with one eigenmode with a positive real eigenvalue. There are two eigenmodes with eigenvalue zero; one corresponds to varying the mean film thickness, and the other to translational displacement of the travelling wave.
The real part of the most unstable eigenvalue decreases with $\alpha$, until it collides with another eigenvalue while still in the right half plane. These two eigenvalues then form a complex conjugate pair, which eventually crosses the imaginary axis with a finite imaginary part, stabilising the system for $\alpha>0.0434$. This stabilisation occurs via a Hopf bifurcation, and so we would expect to observe small-amplitude limit cycles for $\alpha$ just below the critical value.
However, linear stability alone does not mean that the travelling wave is necessarily an attractor when starting from the uniform state, and indeed initial value calculations starting from the uniform film state, as plotted in Fig.~\ref{fig:Travelling_wave_calculations}(c), do not reach the desired travelling wave when $\alpha=0.05$. However, the system successfully converges to the desired travelling wave when $\alpha=0.1$, and converges more rapidly when $\alpha=0.15$. We note that even when the system is converging to the travelling wave, the solution norms show evidence of decaying oscillations.

\subsection{Non-uniform steady states}

We do not know of any non-uniform steady states to the Benney, weighted-residual or Navier--Stokes equations for flow down a planar, unpatterned wall in the absence of suction; instead structures are swept downstream by the underlying flow.
However, as discussed in \citet{Thompson_steady_suction}, the application of steady non-zero suction gives rise to non-uniform steady states, with their own bifurcation structure and stability properties. Moreover, we can often choose the applied steady suction in order to make a given interface shape into a steady solution of the equations.

In order to include both steady suction and feedback, we use an extension of the controls:
\begin{equation}
\label{Nonuniform_control_scheme}
 F = -\alpha [h(x,t) - H(x)] + S(x).
\end{equation}
Here $\alpha$ is the control parameter, and $S(x)$ is the steady component of $F$ that we are free to specify. If $\alpha=0$, $S(x)$ must have zero mean to prevent growth in fluid mass, and thus to allow steady solutions.

\begin{figure}
\includegraphics[width=0.8\textwidth]{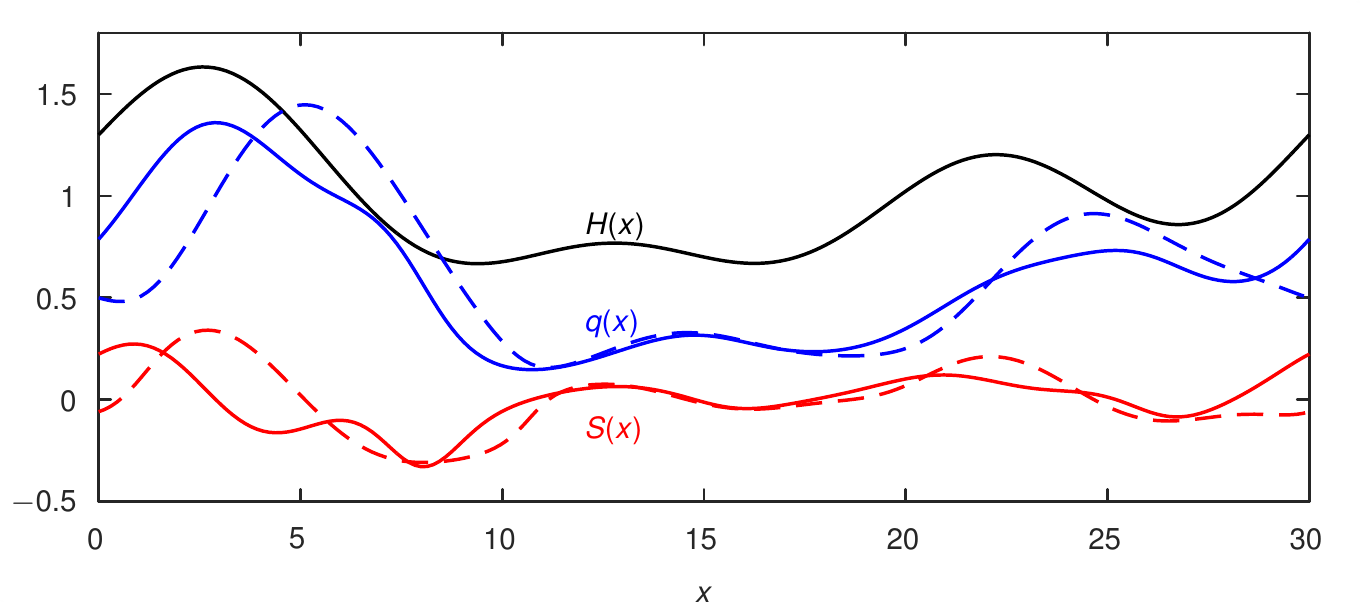}
 \caption{Steady flux $q$ and suction $S$ for the steady state \eqref{nonuniform_sample_state}. The solid and dashed lines correspond to Benney and weighted-residual results, respectively.\label{fig:Compare_steady_S}}
\end{figure}

We choose the following non-uniform steady state as the target state for our calculations:
\begin{equation}
\label{nonuniform_sample_state}
 H(x)  = 1 + 0.3 \cos\left(\frac{2\pi}{L}\right) + 0.2\sin\left(\frac{4\pi}{L}\right)+0.2\sin\left(\frac{6\pi}{L}\right), \quad L=30,
\end{equation}
shown in Fig.~\ref{fig:Compare_steady_S},
and set $R=5$, $C=0.05$, $\theta=\pi/4$. For these parameters, the uniform film state is unstable. The state $h=H$ is not a steady solution of the equations when $S=0$, but we can calculate $S(x)$ to make it so.

For $h=H$ to be a steady solution of the Benney equation, we have
\begin{equation}
 F = S = q_x
\end{equation}
and
\begin{equation}
 q =  \frac{H^3}{3}\left( 2
 -2 H_x\cot\theta + \frac{H_{xxx}}{C}\right)
R \left( 
\frac{8H^6H_x }{15}
-\frac{  2H^4 F}{3}
\right).
\end{equation}
We can rearrange these two equations to obtain a single equation for $S=F$:
\begin{equation}
\label{S_Benney}
 S + \left(\frac{2R H^4 S}{3}\right)_x = \left[\frac{H^3}{3}\left( 2
 -2 H_x\cot\theta + \frac{H_{xxx}}{C}\right)
+
\frac{8 RH^6H_x }{15}\right]_x,
\end{equation}
subject to periodic boundary conditions on $S(x)$.
The right hand side of \eqref{S_Benney} is known, and the left hand side is linear in $S(x)$. 
There is therefore a unique solution for $S(x)$, given $H(x)$, in the Benney model, and the equation has a solution for each smooth, non-zero $H$.
We note that the task of finding a suction profile to enable a particular steady solution is related to inverse topography problems, in which the bottom profile is computed from observations of the interface height \citep{Heining2009} or surface velocity \citep{Heining2013}.

Perhaps unsurprisingly, the linearity with respect to $S$ obtained in \eqref{S_Benney} does not apply in the weighted-residual model. Instead we must solve
\begin{equation}
 F = S = q_x
\end{equation}
and
\begin{equation}
  q =\frac{H^3}{3}\left(2 -2 H_x \cot\theta + \frac{H_{xxx}}{C} \right)
 + R \left( \frac{18q^2 H_x}{35} - \frac{34 H q q_x}{35}+ \frac{ HqF}{5}\right),
\end{equation}
again subject to periodic boundary conditions on $S$ and $q$.
We can use $F=q_x$ to rewrite the second equation as an equation for $q$ alone:
\begin{equation}
  q =\frac{H^3}{3}\left(2 -2 H_x \cot\theta + \frac{H_{xxx}}{C} \right)
 + R \left( \frac{18q^2 H_x}{35} - \frac{27 H q q_x}{35}\right),
\end{equation}
but this is nonlinear in the unknown $q$, and so we cannot guarantee existence or uniqueness of solutions. However, solutions should still exist for $H$ near to $1$, and for the non-uniform state \eqref{nonuniform_sample_state}, we obtain the solution shown in Fig.~\ref{fig:Compare_steady_S}.

With the appropriate $S$ for the corresponding model, as shown in Fig.~\ref{fig:Compare_steady_S}, 
 numerical solutions of the discretised eigenvalue problems described in Sec.~\ref{sec:Eigenvalue_problems} show that the steady state \eqref{nonuniform_sample_state} is stable for $\alpha>1.32$ in the Benney model, and $\alpha>1.39$ in the weighted residual model. In both cases, the exchange of stability occurs via a Hopf bifurcation, so that below the critical value of $\alpha$, we would again expect to observe time-periodic limit cycles.

\begin{figure}
 \includegraphics[width=0.8\textwidth]{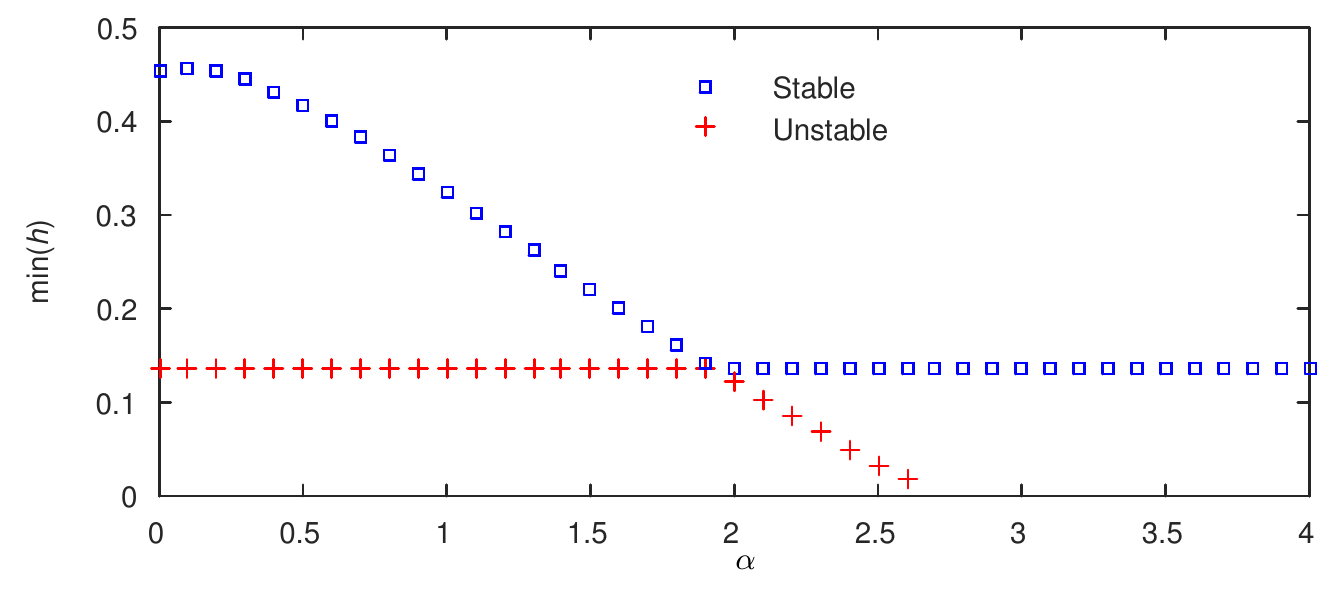}
 \caption{Illustration of a transcritical bifurcation that occurs when controlling to an unstable steady state (with minimum value 0.1375), that has just one positive eigenvalue. Exchange of stability occurs through a transcritical bifurcation at $\alpha=1.92$, necessitating the existence of another solution branch, which here connects to a stable steady solution for the same $S$ at $\alpha=0$.
 The second solution branch only persists slightly beyond the transcritical bifurcation, here diverging through the minimum layer height vanishing at a finite value of $\alpha$. 
 The parameters here are $R=0$, $C=0.05$, $\theta=\pi/4$, $S = 0.7\cos(2\pi x/10)$, which matches the bifurcation structure for $\alpha=0$ shown in Fig. 3 of \citet{Thompson_steady_suction}.
 \label{fig:Transcritical_bifurcation}
 }
\end{figure}

A second mechanism for exchange of stability involves real eigenvalues passing through zero. In Fig.~\ref{fig:Transcritical_bifurcation}, we choose a steady flux $S(x)$ which is known\citep{Thompson_steady_suction} to correspond to two solutions $H(x)$ when $\alpha=0$, 
one of which ($H_1$) is stable, the other ($H_2$) is unstable with one positive real eigenvalue, and show the results of controlling towards the latter, unstable state, $H_2$. Each steady state at $\alpha=0$ gives rise to a solution branch for $\alpha>0$. 
The target state $H_2$ is always a solution, and is stable for $\alpha>1.92$. For $\alpha<1.92$, $H_2$ has a single eigenvalue with positive real part, and this eigenvalue is exactly zero at $\alpha=1.92$. The exchange of stability via a real eigenvalue passing through zero corresponds to a transcritical bifurcation, and implies the local existence of a second solution branch, which here is the branch that connects back to $H_1$ at $\alpha=0$. 
The second branch diverges as $\alpha$ increases beyond $1.92$, here by the minimum film height tending to zero.

\subsection{Controlling towards non-solutions}
\label{sec:Imperfect_model_knowledge}

 In the previous subsection, we assumed that the target state $H$ is an exact solution of the equations, so that the system will remain at $h=H$ if it ever reaches it, and the main questions surround linear stability, which can be directly modified by linear feedback controls. 
 However, in practice, the target state is highly unlikely to be an exact solution, due to discretisation error, imperfectly known parameters, or, more interestingly for our purposes, discrepancies which arise due to calculating travelling waves or the steady flux $S$ according to a low-order model which only approximates the true system.
 We now investigate robustness to model choice by analysing the behaviour of the system when feedback controls are applied towards a state which is not a solution to the governing equations, and so can never be more than transiently achieved.

\begin{figure}
  \includegraphics[width=0.8\textwidth]{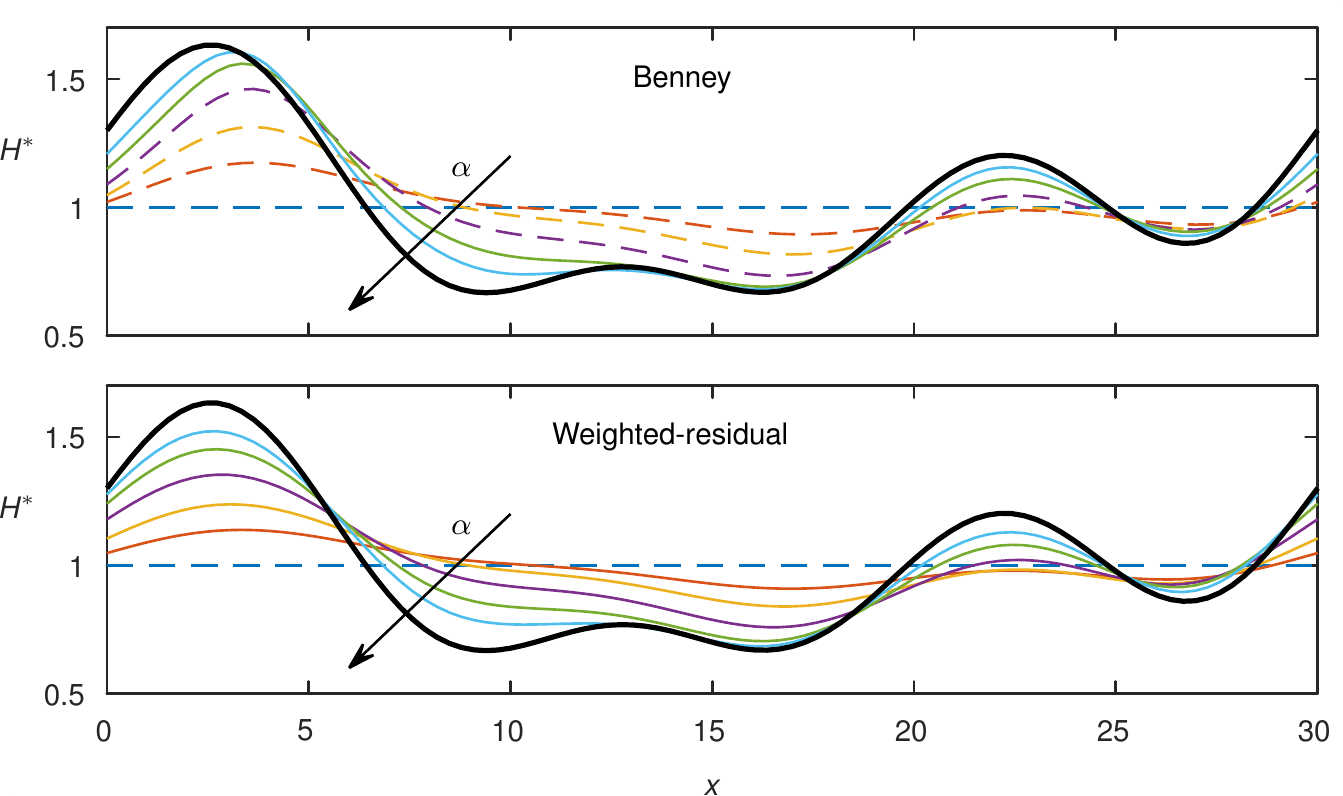}
 \caption{Steady solutions to the Benney equation and weighted residual equations, controlled towards $h=H$ (shown in bold) using the control scheme \eqref{Nonuniform_control_scheme}, for $\alpha = 0, 0.125, 0.25, 0.5, 1, 2$, and $S=0$. Here $R=5$, $C=0.05$ and $\theta=\pi/4$. Dashed lines indicate unstable solutions.
\label{fig:steady_states_alpha_varies}}
\end{figure}

We suppose that the system reaches an equilibrium state $H^*$, which will depend on the target state $H$, the feedback control strength $\alpha$, any patterning imposed on the system via $S$, and the parameters of the uncontrolled system.
We usually have a nonlinear system to solve for $H^*$, which need not have unique solutions. 
In the Benney equation, the steady state $H^*$ must satisfy
\begin{equation}
\label{nonuniform_Benney_mass_control}
 F = -\alpha[H^*-H]+S, \quad F =q_x
\end{equation}
and
\begin{equation}
\label{nonuniform_Benney_flux}
 q =  \frac{H^{*3}}{3}\left( 2
 -2 H^*_x\cot\theta + \frac{H^*_{xxx}}{C}\right)
+
R \left( 
\frac{8H^{*6}H^*_x }{15}
-\frac{  2H^{*4} F}{3}
\right).
\end{equation}
These equations are nonlinear in $H^*$, and can have zero, one, or more solutions.
Fig.~\ref{fig:steady_states_alpha_varies} shows steady solutions to the nonlinear Benney system \eqref{nonuniform_Benney_mass_control}-\eqref{nonuniform_Benney_flux}, and also the corresponding weighted-residual system, for the case $S=0$, with $H$ given by the large-amplitude, non-uniform state \eqref{nonuniform_sample_state}. 
We find that for both models, the numerical solutions for $H^*$ tend towards $H$ as $\alpha$ increases, and our linear stability calculations confirm that the states $H^*$ are stable at large $\alpha$. However, the value of $\alpha$ at which steady states become stable, and also the extent to which the steady states deviate from $H$ at a given $\alpha$, are dependent on the choice of model.

The linearity of the control scheme means that suction can be interpreted as feedback controls towards the equilibrium state $H^*$, which is itself dependent on $\alpha$ and the original target state $H$.
We define $S^*=-\alpha(H^*-H)+S$, so that
for general $h$, we can write
\begin{equation}
 F = -\alpha(h-H)+S = -\alpha(h-H^*) + S^*,
\end{equation}
As a result, the system is indistinguishable from controlling to the state $H^*$ with feedback control parameter $\alpha$ and steady suction component $S^*$.

For large $\alpha$, we can find a simple asymptotic solution for the steady state $H^*$. If the system tends towards a bounded steady state as $\alpha$ increases, then $F$ must remain bounded, and so the interface shape $H^*$ must tend towards $H$. 
Also, $F$ tends towards $S_0(x)$, which is defined to be the steady flux required to make the desired state $H$ a steady solution of the equations.
Thus, without regard to the model details, but assuming only that a bounded steady state $H^*$ exists for large $\alpha$, we find that this state behaves as
\begin{equation}
\label{eq:large_alpha}
 H^* = H + \frac{S-S_0}{\alpha} + O\left(\frac{1}{\alpha^2}\right).
\end{equation}
The function $S_0$, and subsequent terms in the expansion, will depend on the details of the model, but in general we can move the equilibrium state $H^*$ closer to the desired state $H$ by increasing $\alpha$.
In Fig. \ref{fig:Transcritical_bifurcation}, we show a system where there are two steady states for the same parameters, and controls are applied towards one of these states. However, one of the solution branches disappears at a finite $\alpha$, so that for sufficiently large $\alpha$, the only steady state remaining is the one described by \eqref{eq:large_alpha}.
 More generally, branch divergence means that unwanted solution branches can be eliminated by increasing the control amplitude.

\subsection{Control towards non-uniform states with point actuators}
\label{sec:Nonuniform_actuators}

In Sec.~\ref{sec:Actuators_and_observations} we considered control schemes based on localised observers and actuators that remain fixed in the laboratory frame, and showed that these schemes can be used to stabilise the uniform film state. We then showed in Sec.~\ref{sec:Non_uniform} that distributed control schemes can be used to stabilise non-uniform travelling waves, and to create and stabilise non-uniform steady states. 
However, the extension of the point-actuator control schemes to non-uniform travelling waves and non-uniform steady states faces significant difficulties.

Travelling waves are steady with respect to a moving coordinate $\zeta = x-Ut$, and can be written as $h = H(\zeta)$. However, if the observers and actuators are fixed in the laboratory frame, then these move relative to the travelling wave to be controlled. To calculate linear stability, we first transform to the moving frame, so that the base state $h = H(\zeta)$ is a steady solution of the controlled equations. However, the evolution of small perturbations is subject to the spatial structure of the control scheme, which in this moving frame is also time-dependent. If the control scheme is spatially periodic in the laboratory frame, then it is both spatially and temporally periodic in the travelling frame, and we must use Floquet multipliers with respect to time to obtain eigenvectors. For a general control scheme, this requires the computation of eigenfunctions that are explicitly dependent on both space and time within periodic boundary conditions, which is beyond the scope of the present study. We note however that \citet{Susana_long} showed that for the KS equation, localised controls derived for a uniform state could be used to stabilise travelling wave solutions in cases where the wave frame moves relative to the controls.

Non-uniform steady interface shapes $H(x)$ require a non-zero suction profile $S_0(x)$ in order to be steady solutions of the governing equations. However, if suction must be delivered through a linear combination of $M$ localised actuator shapes, it is very unlikely that the exact profile $S_0(x)$ can be achieved. Thus we will no longer obtain the result that $h\rightarrow H(x)$ when strong controls are applied. It is easy to imagine situations where the interface shape appears to be close to the desired state when viewed through localised observers, while diverging significantly at other positions, and so we leave the analysis of this system to future work.

\section{Conclusion}
\label{sec:Conclusion}

In this paper, we analysed the effect of feedback control on the dynamics of a thin film flowing down an inclined plane. Feedback was applied through injection and suction through the planar wall, with the required injection/suction profile determined in response to observations of the position of the air-fluid interface.
We used long wave models based on the Benney and weighted-residual methodologies to describe the effect of suction and injection on the system dynamics.
We note that suction is the only mechanism by which the net system mass can be modified, and so suction controls are the only way in which perturbations of infinite wavelength can be made better than neutrally stable.

The simplest control scheme is to suppose that the suction profile is locally proportional to the deviation of the interface profile from the desired state, so that fluid is injected where the film is particularly thin, and removed from thicker regions. We used a linear stability analysis to show that this simple control scheme, governed only by the constant of proportionality $\alpha$, has a stabilising effect on the uniform film state for positive $\alpha$ in both Benney and weighted-residual models, and also in the Navier--Stokes equations. We calculated the critical value of $\alpha$ needed to stabilise the uniform state to perturbations of all wavelengths, and showed that the control scheme can significantly increase the Reynolds number for the onset of instability.

The analysis summarised so far is for distributed controls, but we also studied a more realistic scenario by supposing that injection/suction can be delivered only via a small number of localised actuators, corresponding for example to slots in the planar wall. Likewise, we should base our control scheme on a limited number of observations of the system state.
The control system requiring the least amount of communication between actuators and observers occurs when each actuator is connected to only one observer, and the applied suction is proportional to the deviation of the observation from the desired value. For equally spaced actuators, our numerical calculations show that this singly-connected control scheme has a stabilising effect on the uniform film state in both long-wave models. The uniform state becomes more stable as the number of observers and actuators is increased. 
We investigated the effect of displacing the observer relative to its linked actuator, and found that the observer should ideally be positioned slightly upstream of the actuator to obtain the best stabilisation. Displacement between observers and actuators can also be incorporated in the fully distributed case, and we again find that the most efficient stabilisation occurs when the observer is slightly upstream of the actuator.

In principle, we should be able to obtain better system performance by using all available observations to compute the feedback controls.
If the entire system state is observable, we can use standard algorithms from control theory to decide the control inputs according to various objectives.
For example, we used the linear quadratic regulator (LQR) algorithm to minimise a cost functional defined in terms of the deviation of the film from the flat state and the actuator amplitudes from zero. 
 The use of point actuators means the system is not translationally invariant, and so the linear stability of the Navier--Stokes equations can no longer be studied by a normal mode analysis.
However, we found that controls calculated using the LQR algorithm for the Benney equation were able to stabilise the uniform state in both the Benney and weighted-residual systems.

For the case where only a small number of observations are available, controls developed under the assumption of full observations can still be implemented by using dynamical observers, and we exploited this strategy to control the Benney system. In this scheme, the Benney system is augmented by a system of ordinary differential equations to create an evolving approximation of the magnitudes of the unstable eigenmodes, which evolves according to our understanding of the underlying system, with corrections due to the available observations. Our stability and initial value calculations confirm that this approach does indeed stabilise the uniform state in the Benney system. For our test case, we found that increasing the number of observations above the number of unstable modes initially yields a significant increase in the overall convergence rate, but further increases have negligible effect.

To test the robustness of the dynamical observer scheme in a proxy physical setting, we inserted the Benney control scheme into the weighted-residual equation. We found that the uniform state was sometimes stable, but this depended sensitively and non-monotonically on the number of observations used to calculate the controls. The eigenvalues of the Benney and weighted residual equations behave differently, and so we might expect that the dynamic approximations converge poorly to the true state.
However, at least for stabilising the uniform state, we have the option of using the singly-connected control scheme with discrete actuators and observers, which behaves similarly in both long-wave models, and so depends relatively weakly on model details.

The thin-film systems can support non-uniform travelling waves, which propagate down the slope at constant speed. These may be stable or unstable; and we found that the locally proportional controls can be used to stabilise unstable travelling waves. The total magnitude of the imposed suction will vanish as the target state is approached if it is an exact solution of the equations, so controls can be in principle be used to physically verify the shape of unstable states.
If a steady suction profile is applied, the system can support non-uniform steady states \citep{Thompson_steady_suction}. These steady states have their own bifurcation structure, can be stable or unstable, and have a more complicated internal flow than that for a film of uniform thickness. If the suction profile corresponding to a desired steady interface shape is known exactly, we showed that the feedback control scheme can be used to stabilise the steady state in a similar manner to that for stabilising travelling waves.

The shape and speed of travelling waves, and the suction profile corresponding to steady states, differs between the two long-wave models here, and likely also the Navier--Stokes equations. It is therefore unreasonable to assume that the target state is an exact solution of the equations. However, we find that if controls are applied with large positive $\alpha$ towards an arbitrary state, the system will both move towards that state and become stable as $\alpha$ is increased, irrespective of the model used.

In some ways it is unsurprising that simple control schemes can be used to linearly stabilise the uniform state in the Benney equation, as the linear operator is similar to that for the Kuramoto-Sivashinsky (KS) equation, where control schemes have been rigorously derived \citep{Susana_long,Susana}. However, the KS results provide no guarantee on the nonlinear behaviour, or on system dynamics away from long-wave limits, and so our nonlinear initial value calculations and linear stability calculations in the Navier--Stokes equations provide meaningful tests on the use of feedback control. It would be interesting to investigate the effect of linear suction controls on nonlinear stability and blow up phenomena in the Benney equation.

The motivation for this work was to act as a first step towards experimental implementation of feedback controls in thin film flow, and indeed the results are very promising, but there are a number of physically-motivated questions still to be addressed.
The analysis in this paper concerns only 2-D flows, but in 3-D simulations or experiments there are more choices regarding the configurations of actuators and observers which may be arranged as points or slots, and we would also need to consider edge effects.
In Sec.~\ref{sec:Control_to_uniform}, we analysed linear stability of a uniform film within an infinite domain, while we later applied periodic boundary conditions for our analysis of control strategies using localised actuators and observers, and for control towards non-uniform states. However, experiments are actually performed on a wall of finite length, with an inlet (at which perturbations can be applied) and outlet. Future work could include exploring the control strategies described in this paper with more realistic boundary conditions, assessing the effect of noise, and also incorporation of restrictions on the control scheme to reflect latency in flow visualization, data processing, and the actual physical setup by which suction and injection is applied.

 \begin{acknowledgments}
We acknowledge financial support from Imperial College through a Roth PhD studentship, and the Engineering and
Physical Sciences Research Council of the UK through grants no. EP/K041134/1, EP/J009636, EP/L020564, EP/L025159/1 and EP/L024926.
We thank Prof. Serafim Kalliadasis and Dr Marc Pradas for helpful discussions.
\end{acknowledgments}

\appendix

\section{Controllability and detectability}
\label{appB}

Here we state some basic definitions from control theory; further details can be found in \citet{Zabczyk1992}.
We consider the linear system
\begin{equation}
\label{eq:def1}
\dot{z} = Az + Bu, \quad y = Cz,
\end{equation}
where $A$, $B$ and $C$ are $N\times N$, $N \times M$ and $M\times P$ matrices, respectively. We will say that a matrix $A$ is stable if all its eigenvalues have negative real part.

We will call the system \eqref{eq:def1}, or the pair $(A,B)$, controllable if there exists a matrix $K$ such that $A+BK$ is stable. If the system is controllable, we can always obtain the state $z^*$ by taking $u = K(z-z^*)$, regardless of initial conditions.
Similarly, we say that system \eqref{eq:def1}, or the pair $(A,C)$, is detectable if there exists a matrix $L$ such that $A + LC$ is stable.
If the pair $(A,C)$ is detectable, then $(A^T,C^T)$ is controllable. 

The Kalman Rank condition gives a necessary and sufficient condition on $A$ and $B$ for controllability, and therefore detectability. This condition states that the system \eqref{eq:def1} is controllable if and only if $\operatorname{rank} [A|B] = N$, where 
\[
[A|B] = [B \,\, AB \,\, A^2B \,\, \cdots \,\, A^{N-1}B]
\]
is a $N\times N^2$ matrix obtained by writing consecutively the columns of the matrices $A^{n-1}B$, $n=1,\dots,N$.

The natural choice when constructing controls based on the observations $y$ would be to choose a matrix $K$ such that the matrix $A + BKC$ is stable. Controls that can be written in the form $u = Ky$ are called \emph{static output feedback controls}. However, for nontrivial $B$ and $C$, it is not possible, in general, to construct a matrix $K$ so that $A+BKC$ is stable.
This difficulty motivates the construction of the \emph{dynamical observers} presented in Sec.~\ref{sec:Evolving_controls}.

\bibliography{Bibliography}

\end{document}